\documentclass[12pt, letterpaper]{article}
\usepackage[margin=2cm]{geometry}

\usepackage{amssymb}
\usepackage{amsmath}
\usepackage{graphicx}

\usepackage{mathtools}

\usepackage{cite}

\newcommand{\dd}{\mathrm{d}}
\newcommand{\dv}[2]{\frac{\dd #1}{\dd #2}}
\newcommand{\qq}[1]{\quad \text{#1} \quad}

\usepackage{nicefrac}   
\newcommand{\nfrac}[2]{\nicefrac{#1}{#2}}

\usepackage{color}
\newcommand{\change}[1]{#1}

\usepackage[ruled,vlined]{algorithm2e}
\newcommand{\pluseq}{\mathrel{+}=}
\newcommand{\compeq}{\mathrel{=}=}

\newcommand{\mat}[1]{\mathbf{#1}}
\newcommand{\matth}[1]{\mathbf{#1}_\theta}

\newcommand{\traj}{\mathcal{T}}

\title{Avoiding matrix exponentials for large transition rate matrices}
\author{Pedro Pessoa$^{1,2}$,Max Schweiger$^{1,2}$,Steve Press\'e$^{1,2,3}$ \\
$^1$Center for Biological Physics, Arizona State University,
Tempe, AZ, USA\\
$^2$Department of Physics, Arizona State University,
Tempe, AZ, USA\\
$^3$School of Molecular Sciences, Arizona State University,
Tempe, AZ, USA}
\date{}

\begin{document}

\maketitle 

\begin{abstract}
 Exact methods for exponentiation of matrices of dimension $N$ can be computationally expensive in terms of  execution time ($N^{3}$) and memory requirements ($N^{2}$) not to mention numerical precision issues. A matrix often exponentiated in the natural sciences is the rate matrix. Here we explore five methods to exponentiate rate matrices some of which apply more broadly to other matrix types. Three of the methods leverage a mathematical analogy between computing matrix elements of a matrix exponential and computing transition probabilities of a dynamical processes (technically a Markov jump process, MJP, typically simulated using Gillespie). In doing so, we identify a novel MJP-based method relying on restricting the number of ``trajectory" jumps  \change{that incurs improved} computational scaling. 
 We then discuss this method's downstream implications on mixing properties of Monte Carlo posterior samplers. We also benchmark two other methods of matrix exponentiation valid for any matrix (beyond rate matrices and, more generally, positive definite matrices) related to solving differential equations: Runge-Kutta integrators and Krylov subspace methods. Under conditions where both the largest matrix element and the number of non-vanishing elements scale linearly with $N$ --- reasonable conditions for rate matrices often exponentiated --- computational time scaling with the most competitive methods (Krylov and one of the MJP-based methods) reduces to $N^2$ with total memory requirements of $N$. 

\end{abstract}

% Body of paper goes here. Use proper sectioning commands. 
% References should be done using the \cite, \ref, and \label commands

%%%% ============================================== %%%%
%%% Here the main text starts

\newpage
\tableofcontents
\newpage

\section{Introduction}

Matrix exponentiation is a common operation  across scientific computing. As a concrete example of how matrix exponentials arise, we consider a system evolving through a discrete state space whose probability of occupying distinct states follows a linear differential equation of the form \cite{Ross,VanKampen,Presse23, Lee12}
\begin{equation}\label{differential_ME}
    \dv{\bar{\rho}(t)}{t} = \bar{\rho}(t)\matth{A} \ ,
\end{equation}
where $\bar{\rho}(t)$ is a  {(row)} vector of probabilities for the system's states and $\matth{A}$ is the generator matrix also called a transition rate matrix  {(where all rows sum to one)}.  
\change{Usually, the elements of $\matth{A}$ are functions of small number of kinetic parameters -- \emph{e.g.}, the propensity of each chemical reaction modelled. We denote this concise set of unique parameters as $\theta$. To illustrate this with practical applications, Sec. \ref{sec:examples} provides specific examples. }

\begin{figure}[p]
\begin{center}
    \includegraphics[width=\textwidth]{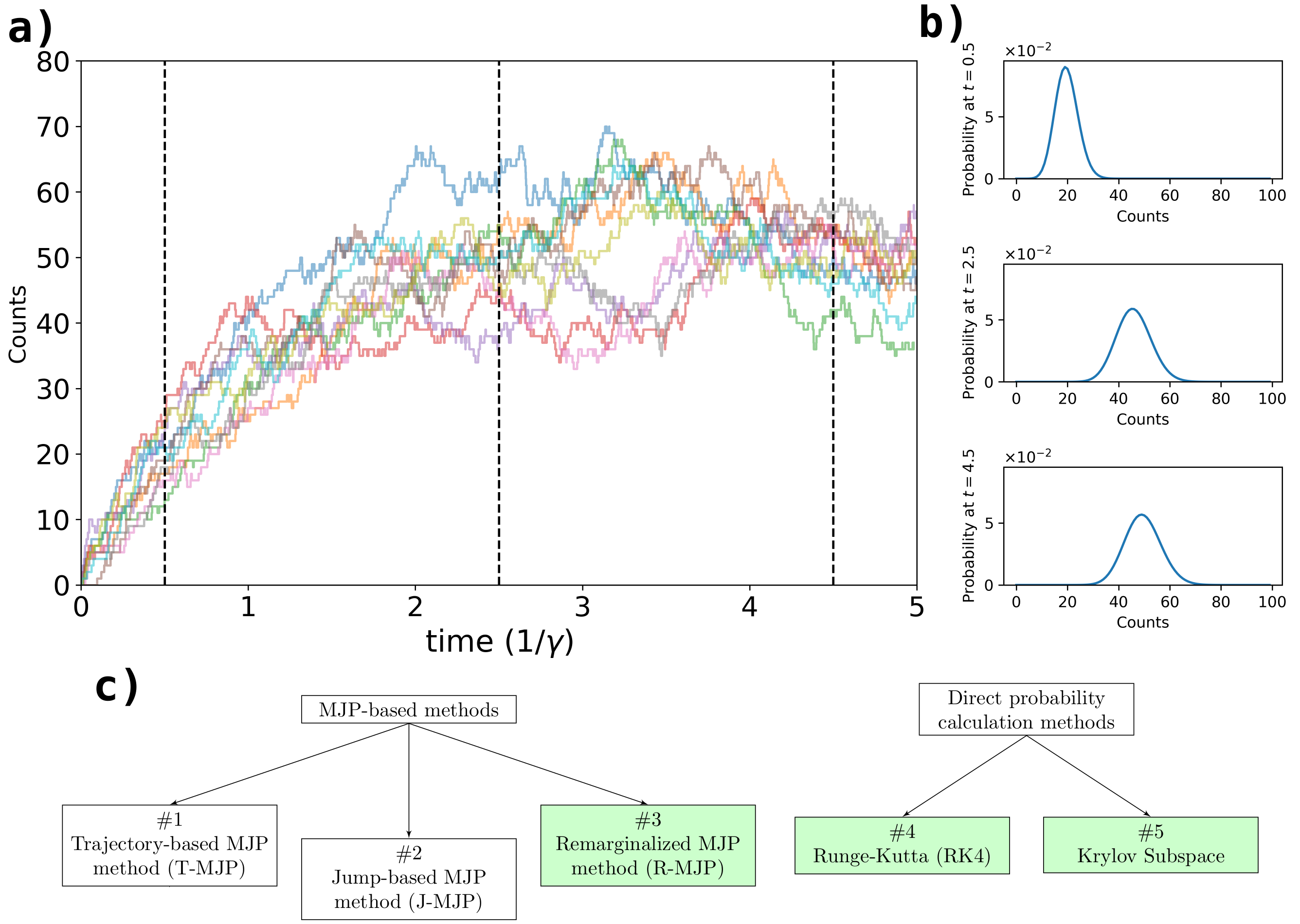}
\end{center}
\caption{
Diagram depicting MJP trajectories, probability distributions, and the methods included in our benchmark. In {\bf a)}, we present sample trajectories of an MJP model of a birth-death process, detailed in Sec. \ref{sec:Birth-Death}. In {\bf b)}, three plots represent the probability distributions over states at times indicated by the dashed lines in {\bf a)}. In {\bf c)}, methods \#1 --- \#5 are categorized. 
Methods with the best scaling are highlighted in green. 
For more details, refer to Sec. \ref{sec:inference}. }\label{fig:intro-cartoon}
\end{figure}

Systems described by \eqref{differential_ME} capture diverse phenomena such as allosteric enzyme control \cite{hung2022allosteric}, chromatin reorganization \cite{fletcher2022non}, 
metabolic interactions \cite{Presse10,rios2007experimental}, transcriptional regulation \cite{Kilic23,Munsky18,Tiberi18,Presse11}, 
and tumor growth \cite{gatto2020pan}. 

Modeling systems described by \eqref{differential_ME}  typically requires integrating \eqref{differential_ME}, from time $0$ to time $t$, yielding
\begin{equation}\label{ME_solution}
    \bar{\rho}(t) = \bar{\rho}(0) e^{\matth{A}t} \ ,
\end{equation}
where $\bar{\rho}(0)$ is a vector of state probabilities at time $0$. 

As \eqref{differential_ME} is often invoked to describe complicated physical systems, $\matth{A}$'s dimensionality (physically interpreted as the number of states \change{the system can occupy}), $N$, quickly grows, and direct matrix exponentiation becomes unwieldy for two reasons: 
\textbf{(i)} even as a straightforward operation in linear algebra, the dominant exact matrix exponential method's cost scales as $N^3$. 
Although approximations (\emph{e.g.}, the Padé approximation and the ``scaling and squaring'' algorithm \cite{Moler03}) reduce absolute computational times, they rely on matrix diagonalization or inversion also scaling as $N^3$, \change{as outlined in Ref.\cite{Moler03}}; \textbf{(ii)} for dynamics assumed irreducible, {\it i.e.}, when it is possible to go from any state to any other state in finite time with non-zero probability, the exponential \change{$ e^{\matth{A}t}$} is fully dense and the memory required to store the matrix exponential scales with $N^2$. To give some idea of the concrete limitations posed by \textbf{(i)} and \textbf{(ii)}, when $N$ is only $44722$, the theoretical minimum memory allocation for a dense square matrix of double precision floats exceeds $16\text{GB}$ (a common memory allocation for multi-core computing systems as of this writing). 

There are redeeming features of scientific computing that help alleviate the worst-case-scenario $N^2$ memory scaling of storing the generator matrix $\matth{A}$. 

In fact, in modeling inspired from chemical systems, whose acceptable state spaces vastly exceed the number of reactions allowed, we often encounter sparse generator matrices. Indeed, properly exploiting sparse matrices provides an immediate cost benefit since storing them requires memory scaling with the number of non-zero elements, \emph{i.e.}, \change{linear scaling or better with $N$}. Furthermore, \change{since for irreducible dynamics the matrix exponential will be fully dense,} avoiding dense matrix exponentiation using only sparse matrix-vector products incurs an additional benefit to computational time scaling, reducing the total computational cost to scale with the number of non-zero elements, \change{which %often \textcolor{blue}{'often' here seems to contradict above... it's either linear or better or it isn't; which one?} 
even beyond modeling chemical systems, often scale linearly or better with $N$. }

The benefit of using algorithms tailored to sparse matrices is particularly important since the most common statistical inference methods --- that is, iterative methods such as Markov chain Monte Carlo (MCMC)  --- require repeated calculation of likelihoods obtained by taking the product of many probabilities of the form of \eqref{ME_solution} at each Monte Carlo iteration \cite{Presse10, Jazani19, BryanIV22, Sgouralis18, Munsky18, Sukys22, Kilic23, saurabh2023single, saurabh2023singleII,safar2022single,kilic2023monte}, augmenting the already challenging task of exponentiating a fixed matrix for systems of large $N$ \cite{Munsky06, Sidje15, Vo17, Gupta17, Sukys22}. Taken together, these considerations raise the question: how can we avoid time \textbf{(i)} and memory \textbf{(ii)} problems involved in matrix exponentiation under a sparse matrix assumption?

To determine the best available solution to problems \textbf{(i)} and \textbf{(ii)}, we identify five alternative methods, organized in Fig. \ref{fig:intro-cartoon} {\bf c)}, elaborated in detail in Sec. \ref{sec:inference}. 
In particular, the first three methods rely on the mathematics originally formulated to simulate MJPs \cite{Grassmann77} and subsequently adapted to dynamical inference \cite{Rao13,Kilic21generalizing,Zhang10}. We refer to the three methods leveraging the MJP structure underlying \eqref{differential_ME} as MJP-based methods.

The first MJP-based method (\#1), which we refer to as trajectory MJP-based method (T-MJP) was developed by Rao and Teh in Ref. \cite{Rao13}. In T-MJP the system's trajectory is sampled through the forward-filtering backward-sampling algorithm \cite{Rao13,Carvalho10,Jazani19,Presse23, Kilic21generalizing}. This method learns the parameters $\theta$ in a manner that is statistically equivalent to exactly solving \eqref{differential_ME}. As we will show, T-MJP's 
trajectory sampling (as a substitute for matrix exponentiation) time scaling is $N^2$ (an improvement over $N^3$ for naive matrix exponential), but unfortunately retains the $N^2$ memory scaling described in \textbf{(ii)}. Even worse, since T-MJP avoids the matrix exponential by introducing a large number of latent random variables ({\it i.e.}, the trajectories which includes the jump times and states visited after each jump), it severely increases the dimensionality of any statistical inference task. As a result, many MCMC iterations are needed to sample both trajectories and parameters. That is, the MCMC chains suffer from ``poor mixing" giving rise to slow convergence.
These computational issues may be avoided if trajectory samples are not needed. 

Method \#2 avoids the matrix exponential by sampling only the number of discrete state jumps (details in Sec. \ref{sec:forward}), both reducing the memory cost scaling to $N$ and improving mixing. Since it samples only jumps, we refer to this method as the jump MJP-based method (J-MJP). As we will see later, although J-MJP exhibits marginally better mixing than T-MJP, it still introduces a large number of latent variables (the number of jumps per trajectory).

Method \#3, referred to as remarginalized MJP-based method (R-MJP), improves mixing by entirely marginalizing over trajectories, \change{meaning integrating over all trajectories ending in the same observation}. Although marginalizing over trajectories demands an approximation not required by Methods \#1 and \#2, R-MJP provides an upper bound in the total error \change{in the final result},
with no additional computation \cite{Zhang10}. With the same scaling as both previous methods, we will demonstrate R-MJP's considerably better mixing than Methods \#1 and \#2.

The two remaining methods treat the solution of \eqref{differential_ME} without leveraging trajectories, or indeed any facet of MJP structure. We therefore refer to them as direct probability calculation methods. 

Method \#4 simply applies the Runge-Kutta method for solving differential equations \cite{Butcher00} to avoid matrix exponentiation. Method \#4 proceeds by slicing the time projection into small intervals of size $\Delta t$ and evolving \eqref{differential_ME} within that time interval. Naturally the choice of $\Delta t$ is arbitrary, making it significantly harder to keep track of the upper bound in the approximation error. 
Thus $\bar{\rho}(t)$ arises from the sequence $\{ \bar{\rho(0)}, \bar{\rho(\Delta t)},\bar{\rho(2 \Delta t)},$ $ \ldots , $ $\bar{\rho(t)} \}$. We focus in particular on the fourth-order Runge-Kutta methods (RK4), an implementation typically chosen as a balance between precision and computational time\cite{Butcher00}. As usual with RK4, time scaling is $N^2$ and memory requirements scale as $N$.

The final method (\#5) is the Krylov subspace approximation of \eqref{differential_ME} \cite{gaudreault18,Vo17,Vo19} discretizing time in the same manner as RK4. It approximates the matrix exponential by generating a Krylov subspace of $\kappa$ vectors as $\{\bar{\rho}, \bar{\rho} \matth{A}, \bar{\rho} \matth{A}^2,$
$ \ldots , \bar{\rho} \matth{A}^{\kappa-1}\}$ 
and exponentiating the matrix projection of $\matth{A} \Delta t$ into the subspace. Here problems \textbf{(i)} and \textbf{(ii)} associated with calculating the exponential of $\matth{A}$, are avoided by calculating the exponential of the smaller $\kappa \times \kappa$ projection, rather than the $N\times N$ matrix exponential. 

In what follows, we implement methods \#1-\#5 comparing their computational time scaling, memory requirements, and mixing.  \change{We do so in order to determine which method is best suited for iterative MCMC aimed at estimating parameters and their associated uncertainty. } 
Due to method \#1's excessive memory requirements, we avoid comparing it to other methods and proceed by benchmarking the remaining methods (\#2-\#5). 
We show that when compared to RK4 and R-MJP, J-MJP requires a much larger number of matrix exponential equivalent calculations to complete a modeling task, as predicted.  From our benchmark, we conclude that either R-MJP (Method \#3) or Krylov subspace (Method \#5) attains the fastest overall computational time, depending on both the state space and the dynamics at hand. \change{A guide for when to use each method is given in the conclusion 
(Sec. \ref{sec:conclusion})}. A diagram illustrating the methods and the reasons some are considered unsuitable are shown in Fig. \ref{fig:intro-cartoon} {\bf c)}.

\section{Methods}

\change{
This section outlines two key points: first, as an overview, we introduce the necessary background on Markovian dynamics within discrete state spaces in Sec. \ref{background}. Next, in Sec. \ref{sec:inference}, we elaborate on the key aspects required in implementing Methods \# 1 -- \# 5. While technical details presented herein are essential for those intending to immediately apply these methods, readers primarily interested in application and benchmarking may directly proceed to Sec. \ref{sec:results}. Moreover, code for the implementation of methods \#2 -- \#5 (using the Python library for compiled sparse matrices \cite{smn}) can be found in our GitHub repository \cite{github}.
}

\subsection{Background on MJPs}\label{background}

Here we describe a general theory of continuous time systems transitioning between distinct states, $\sigma_n$ \cite{Ross,VanKampen,Presse23,Kilic21generalizing}. We collect any system's states into a set termed the state space $\sigma_{1:N} = \{\sigma_0, \sigma_1, \ldots, \sigma_{N-1}\}$ 
We describe the transition dynamics within the state space by a transition rate matrix $\mat{\Lambda}$ whose elements $\lambda_{nm}$, called transition rates from state $\sigma_n$ to $\sigma_m$, imply that the probability of the system transitioning from $\sigma_n$ to $\sigma_m$ in an infinitesimal time interval $\dd t$ is $\lambda_{nm} \dd t$. By definition, all self-transition rates are zero, $\lambda_{nn} = 0 \ \forall n$, leaving $N(N-1)$ potentially independent parameters. 

In many problems of scientific interest, \change{of which some examples are provided} in Sec. \ref{sec:examples}, we will see that the remaining $\lambda_{nm}$ are typically described using a much smaller number of unique parameters, collected under $\theta$, from which $\lambda_{nm}$ are \change{determined}. For example, in a simple death process, the rates in going from $n$ members to $n-1$ members, $\lambda_{n, n-1}$, are all the same irrespective of $n$ (and coincide with the death rate). To make $\theta$ dependence explicit for matrices, we define $\mat{\Lambda}_\theta = \mat{\Lambda}(\theta)$ and denote the count of non-zero elements in a matrix $\mat{M}$ as $[\mat{M}]$. {For example, for a simple death process starting from $N$ members, we have $N$ non-zero elements.}

Such systems in discrete space evolving in continuous time with rate matrices are termed MJPs. An MJP's dynamics in a time interval $[0,T]$, usually thought of as an experimental time-course, are described completely by the states the system occupies over time, $\Tilde{s}_0, \Tilde{s}_1, \ldots, \Tilde{s}_l$ \change{with each of these occupied states $\Tilde{s}_{l'}$ being one of the state space elements, $\Tilde{s}_{l'} \in  \sigma_{1:N} $}, and the time at which each transition occurs, $t_1, t_2, \ldots, t_l$. Together, these describe a trajectory, an equivalent description of the MJP's dynamics as function $\traj(\cdot)$ 
mapping times from the interval $[0,T]$ onto the state space 

\begin{equation}\label{defining_trajectory}
    \traj(t) = 
\begin{cases} 
\tilde{s}_0 & \text{if } 0\leq t < t_1 \\
\tilde{s}_{l'} & \text{if } t_{l'} \leq t < t_{l+1} \text{ and } l' \in \{1,2,\ldots, l-1 \} \\
\tilde{s}_l & \text{if } t_l\leq t < T
\end{cases} \  .
\end{equation}
Examples of many superposed trajectories are plotted in Fig. \ref{fig:intro-cartoon} {\bf a)}.

\subsubsection{Forward Modeling}\label{sec:forward}

For now, we are interested in describing the probabilities over states, $p(\traj(t) = \sigma_n|\theta)$, occupied over this interval $[0,T]$.
To do so, we begin by concretely defining $\bar{\rho}$ and $\matth{A}$ in \eqref{differential_ME}. 
We define $\bar{\rho}(t) = (\rho_0(t),  \rho_1(t),   \ldots, $ $  \rho_{N-1}(t))$ as a probability vector, with each element $\rho_n(t)$ denoting the probability that the system is in state $\sigma_n$ at time $t$, 
\begin{equation}\label{vector_definition}
    \rho_n(t) \doteq p(\traj(t) = \sigma_n|\theta)  \ .
\end{equation} 
The rate matrix, $\mat{\Lambda}$, described at the beginning of Sec. \ref{background} is closely related to the generator matrix, $\matth{A}$. First appearing in \eqref{differential_ME}, $\matth{A}$'s elements $a_{nm}(\theta)$ are
\begin{equation}\label{A_def}
    a_{nm}(\theta) \doteq 
    \begin{cases}
    \lambda_{nm}(\theta)  \quad & \text{if} \quad n \neq m \\ 
    -\sum_\nu\lambda_{n\nu}(\theta)  \quad & \text{if} \quad n = m 
    \end{cases} \ .
\end{equation}
Importantly, as all $\lambda_{nm}$ are non negative, ${|a_{nn}|} \geq |a_{nm}| \; \forall m$, then the largest element of $\|\matth{A}\|$ is along the diagonal. Additionally, as $\lambda_{nn}=0 \; \forall n$, $\matth{A}$ has more non-vanishing elements. To wit, $[\matth{A}] = [\matth{\Lambda}]+N$.  With our notation established, we can describe the evolution of $\bar{\rho}(t)$ by \eqref{differential_ME}.

To compute $\bar{\rho}(t)$ leveraging the mathematics of MJPs, we define a matrix we call the uniformized transition matrix
\begin{equation}\label{B_def}
    \matth{B} \doteq \matth{I}+\frac{\matth{A}}{\Omega_\theta} \ ,
\end{equation}
where $\Omega_\theta$ is an arbitrary number greater than the magnitude of any one element of $\matth{A}$, such that all elements of $\matth{B}$ are positive. Although it can take any value larger than $\max\limits_{n}{|a_{nn}|}$, we choose $\Omega_\theta$ as $\Omega_\theta = 1.1 \max\limits_{n}{|a_{nn}|}$. Note that $\matth{A}$ and $\matth{B}$ have the same number of non-zero elements ($[\matth{A}] = [\matth{B}]$), and, from \eqref{A_def}, $\matth{B}$'s rows sum to one.
Using \eqref{B_def}, we can rewrite \eqref{ME_solution} as
\begin{equation}\label{ME_expansion}
\begin{split}
    \bar{\rho}(t) 
    & = \bar{\rho}(0) e^{\matth{A}t} = \bar{\rho}(0) e^{\Omega_\theta \matth{B}t}   e^{-\Omega_\theta t}   \\
    & = \bar{\rho}(0) \left( \sum_{k=0}^\infty \frac{(\Omega_\theta t)^k \matth{B}^k}{k!}  \right) e^{-\Omega_\theta t} \\
    & = \sum_{k=0}^\infty \bar{\rho}(0) \matth{B}^k \left( \frac{(\Omega_\theta t)^k}{k!}   e^{-\Omega_\theta t}  \right) .
\end{split}
\end{equation}
We recognize the Poisson density within brackets, $\text{Poisson}(k|\Omega_\theta t)$,
\begin{equation}\label{ME_uniform}
    {\rho}_n (t) = p(\traj(t) = \sigma_n|\theta) =  \sum_{k=0}^\infty (\bar{\rho}(0) \matth{B}^k)_n \  \text{Poisson}(k|\Omega_\theta t) \ , 
\end{equation}
and interpret \eqref{ME_uniform} as a marginalizing $k$ out of
$p(\traj(t)=\sigma_n|k,\theta)$ such that
\begin{equation}\label{ME_uniform_conditional}
    p(\traj(t) = \sigma_n|\theta) =  \sum_{k=0}^\infty p(\traj(t)=\sigma_n|k,\theta)  \  p(k|\theta)  \ ,
\end{equation}
leaving
\begin{equation}\label{final_cond}
    p(\traj(t)=\sigma_n|k,\theta) = (\bar{\rho}(0) \matth{B}^k)_n \qq{and}  p(k|\theta) = \text{Poisson}(k|\Omega_\theta t) \ . 
\end{equation}
Now, sampling trajectories from \eqref{final_cond} yields MJP trajectories mediated by $\matth{A}$, described in Sec \ref{background}. As proven in Ref. \cite{Grassmann77}, the jump times themselves may be uniformly sampled across $[0,t]$, resulting in using the term ``uniformization" to describe the  use of \eqref{final_cond} in the simulation of MJPs  \cite{Grassmann77,Rao13,Presse23}. Unlike the Gillespie simulation, $k$  \change{also} includes self-transitions for uniformization to remain equivalent to its corresponding MJP under uniform jump time statistics. We therefore follow the uniformization literature and say $k$ counts `virtual jumps' (\emph{i.e.} it is distinct from the number of transitions $l$ in the previous section), to indicate that it includes both real and self-transitions. 
Although uniformization is unpopular relative to Gillespie's more efficient stochastic simulation algorithm \cite{Gillespie77}, we describe it here since \change{it is the standard in proposing trajectories conditioned on data subsequently accepted or rejected within Monte Carlo. }

\subsubsection{Inverse Modeling}\label{sec:Bayes}

With the details of forward modeling established, we turn to the task, the so-called inverse problem, \change{requiring} the probability of \change{states} $\rho_n$, \change{given} $\theta$, usually \change{by} matrix exponentiation. In the inverse problem, information on $\theta$ is \change{to be decoded} from observations. Notably, Methods \#1 and \#2 provide stochastic samples of trajectories and jumps, respectively, contributing to the matrix exponential while Methods \#3 -- \#5, in principle, can be used to compute the vector times matrix exponential directly as in \eqref{ME_solution}. 
Later, Sec. \ref{sec:benchmarking}, we will embed all methods within an inverse modeling scheme as a means of assessing the efficiency of each method. 
\change{In order to do so, we immediately} consider multiple back to back intervals, $[0,T]$. We collect the end times of each interval as the set $\bar{T} = \{T^1,T^2, \ldots, T^I\}$. The state of a system could be observed at these times leading to a sequence of state observations $\bar{s} = \{s^1,s^2, \ldots, s^I\}$. Alternately, we could also consider a different \change{scheme involving} a collection of systems at times $\bar{T} = \{T^1,T^2, \ldots, T^I\}$ \change{where each system's state is observed at each of the times} also yielding a sequence of state observations $\bar{s} = \{s^1,s^2, \ldots, s^I\}$. The latter falls within the realm of what is sometimes called snapshot data\change{\cite{Kilic23}}. The methods we discuss here apply to both \change{types} of observations though, for ease of computation, our benchmarking of Methods \#1 -- \#5 is performed on snapshot data.
 
\change{Following notation established in \eqref{defining_trajectory}, from the state observations we write $p(s^i|\theta) = p(\traj^i(T^i) = s^i |\theta)$, often termed the likelihood. }
As defined in \eqref{vector_definition}, we have $p(s^i|\theta) = \rho(T^i)_{s^i} = (\bar{\rho}(0) e^{\matth{A} T^i})_{s^i}$\change{,} equivalent to what is obtained by the vector times matrix exponential in \eqref{ME_solution}. 

In the following section we will explain different ways to avoid the full matrix exponential calculation, but first, we establish the notation necessary to explain Methods \#1 -- \#3.
Learning the kinetic parameters $\theta$ from the data means obtaining $p(\theta|\bar{s})$ \change{related to the} likelihood through Bayes' theorem
\begin{equation}\label{Bayes}
    p(\theta|\bar{s}) = p(\theta) \frac{p(\bar{s}|\theta)}{p(\bar{s})} \propto p(\theta) p(\bar{s}|\theta) \ .
\end{equation}

For Methods \#1 and \#2, we \change{do not directly compute} $p(\theta|\bar{s})$ \change{but rather introduce latent variables} $\traj^i$, the trajectories, and $k^i$, the number of virtual jumps.  That is, we re-write the posterior \change{we desire as arising from} the marginalization over trajectories and virtual jumps, respectively,
\begin{subequations} 
\begin{align}\label{Bayes_others}
     p(\theta|\bar{s}) = \sum\limits_{\overline{\traj}} p(\theta,\overline{\traj}|\bar{s}) 
     &\propto \sum\limits_{\overline{\traj}} p(\theta,\overline{\traj}) p(\bar{s}|\theta,\overline{\traj}) \qq{and} \\
     p(\theta|\bar{s}) = \sum\limits_{\bar{k}} p(\theta,\bar{k}|\bar{s}) 
     &\propto \sum\limits_{\bar{k}} p(\theta,\bar{k}) p(\bar{s}|\theta,\bar{k}) 
     \ .
\end{align}
\end{subequations} 
\change{In other words, we `de-marginalize' the trajectories based on $\bar{k}$.}

In spite of its importance, a closed-form expression for the posterior is often not attainable. In such cases, we can instead draw samples from the posterior using techniques such as MCMC, which constructs a random walk in parameter space whose stationary distribution is the posterior. When we sample the de-marginalized posteriors in  \eqref{Bayes_others}, we marginalize out all but the variables of interest ($\theta$) as a post-processing step. 
In practice, this just means ignoring all sampled values of the latent variables. More details will be given for each Method \#1 -- \#5 below.

\subsection{Avoiding the matrix exponential in Methods \#1 -- \#5}\label{sec:inference}

For each method mentioned in the introduction and described in Fig. \ref{fig:intro-cartoon}, we describe how each method allows one to calculate the state probabilities and discuss cost scaling, summarized in Table~\ref{tab:requirements}. Although each method deals with the time evolution in its own manner, time scaling with $N^2$ appears repeatedly in Table \ref{tab:requirements}. This is because each method requires a loop where $\bar{\rho}$ is multiplied by either $\matth{B}$ (in Methods \#1 -- \#3) or $\matth{A}$ (in Methods \#4 and \#5) sequentially within a single calculation. 
Therefore, we determine the scaling on total time cost using the cost of each loop's iteration and the total number of iterations needed to calculate the full vector times matrix exponential product \eqref{ME_solution}.

We obtain \change{concise time and memory} scaling between different methods under two assumptions inspired by relevant physical examples: a) the number of nonzero elements in $\matth{A}$ scales linearly with $N$ (conveniently guaranteeing sparse $\matth{A}$) and b) the exponent of any power of $N$ appearing explicitly in $\matth{A}$ is less than or equal to $1$ as larger rates can introduce more jumps in the trajectories introduced in Methods \#1 -- \#3 and reduce the time step of integration of Methods \#4 and \#5. 
To be clear, both conditions are used to \change{describe how the time and memory required to calculate likelihood scales with the number of particles. They are not, however,} required for the methods to work.

Regardless of which method we are discussing, when dealing with chemical reactions, the number of nonzero elements scales at worst linearly with the state space size, $O([\matth{A}]) = O(N)$, as in condition a). It is also usually the case that when molecular species do not interact with themselves, the largest element of the rate matrix scales linearly with the state space --- $O(\max\limits_{n}{|a_{nn}|}) = O(N)$, as in condition b). 
As we will see, all methods have some dependence on the largest value in the rate matrix, $\max\limits_{n}{|a_{nn}|}$. As this quantity represents the fastest time-scale at which dynamics occur, all methods must \change{take such scale into account in order to provide accurate results. }

\begin{table}[h]
\centering
\caption{Table summarizing the methods discussed, their number of iterations (first column) and time cost per iteration (second column), with $\kappa$ being the number of vectors taken into the Krylov subspace.
The third column (time scaling) is built by multiplying the first and second columns, assuming conditions a) and b) outlined in the second paragraph of Sec. \ref{sec:inference}. Although J-MJP appears competitive with other methods, we will see in Sec.~\ref{sec:J-MJPac} that the latent variables it requires lead to poor MCMC mixing. \change{The variable $k^\ast$ related to the R-MJP method (\#3) is properly defined in Sec. \ref{sec:BayesR-MJP}, here it is important to know it scales proportionally to $\max\limits_{n}{|a_{nn}|}$.}
}\label{tab:requirements}
\begin{tabular}{l||c|c|c|c|c}
\footnotesize
       & Number                          & Cost per iteration                                & Time scaling                 & Memory required             & Requires             \\
       & of iterations                   &                                                   &                              &             & latent variables             \\
       \hline
T-MJP    & $\max\limits_i  k^i$                                 & $O([\matth{A}])$                                          & $O(N^2)$                     & $N^2+[\matth{A}]$                     & Yes \\
J-MJP    & $\max\limits_i  k^i$                                 & $O([\matth{A}])$                                          & $O(N^2)$                     & $N+[\matth{A}]$                    & Yes \\
R-MJP    & $k^\ast$                                     & $O([\matth{A}])$                                          & $O(N^2)$                     & $N+[\matth{A}]$  & No \\
RK4    & $T (\max\limits_{n}{|a_{nn}|})$              & $O([\matth{A}])$                                         & $O(N^2)$                     & $4N+[\matth{A}]$                   &No \\
Krylov & $\frac{T}{2\kappa}(\max\limits_{n}{|a_{nn}|})$& $O(\kappa[\matth{A}] + \kappa^2 N +\kappa^3)$        & $O(\kappa^2 + \kappa   N^2)$  & $\kappa^2 + \kappa N + [\matth{A}]$ & No\\
\end{tabular}
\end{table}

\subsubsection{Method \#1 -- Trajectory-based MJP method (T-MJP)}

Informed by a set of observed states, $\bar{s}$, we can alternately sample $\overline{\traj}$ and $\theta$ from their respective distributions following a Gibbs sampling scheme \cite{Gelfand90,Presse23}
\begin{subequations} \label{sample_traj_rt}
    \begin{align}
             p(\overline{\traj}|\bar{s}, \theta) \propto  p(\overline{\traj}|\theta) \ p(\bar{s}|\overline{\traj},\theta)  , \quad \text{and} \label{sam_traj_RT}\\
             p(\theta|\overline{\traj},\bar{s}) \propto p(\theta) \ p(\overline{\traj}|\theta) \ p(\bar{s}|\overline{\traj},\theta). \ \label{sam_th_RT}
    \end{align}
\end{subequations}
When enough alternating samples are drawn, we obtain a model of the dynamics (that is, samples of $\theta$ informed by $\bar{s}$) \change{avoiding entirely} the matrix exponential itself.

Sampling $\overline{\traj}$ is detailed elsewhere \cite{Rao13,Kilic21generalizing,Presse23} but its cost lies mainly in the required forward-filtering backward-sampling algorithm \cite{Presse23} within uniformization \cite{Grassmann77}, since it involves storing the intermediate probability distributions (filters) associated with a set of virtual jumps. In the previously defined notation, this means saving the vectors $\{\bar{\rho}(0), $$ \bar{\rho}(0)\matth{B},$$ \bar{\rho}(0)\matth{B}^2,$ $ \ldots,$ $ \bar{\rho}(0)\matth{B}^{k_{\max}}\}$ up to the largest sampled number of virtual jumps $ k_{\max} = \max\limits_i  k^i$. When condition b) is met, this memory requirement scales as $O(N^2)$ ({\it i.e.}, problem \textbf{(ii)}).

\subsubsection{Method \#2 -- Jump-based MJP method (J-MJP)}\label{sec:BayesJ-MJP}

Avoiding both matrix exponentiation and the memory requirements of 
Method \#1, Method \#2 proceeds by alternating between samples of the number of virtual jumps $\bar{k}$ and $\theta$ respectively also following a Gibbs sampling scheme 
\begin{subequations} \label{sample_J-MJP}
    \begin{align}
              p(\bar{k}|\bar{s}, \theta) & \propto  p(\bar{k}|\theta) \ p(\bar{s}|\bar{k},\theta) , \quad \text{and} \label{sam_k}\\
             p(\theta|\bar{k},\bar{s}) & \propto p(\theta) \ p(\bar{k}|\theta) \ p(\bar{s}|\bar{k},\theta) \ . \label{sam_th}
    \end{align}
\end{subequations}
As with the previous method, for a sufficiently large number of alternate samples, we can \change{avoid matrix exponentiation} entirely and still obtain the appropriate probabilities for the states given the kinetic \change{parameters}, $\theta$, and the number of virtual jumps for respective observations, $\bar{k}$.
In Sec. \ref{sec:J-MJPac} we will discuss  practical challenges this poses.

Since each $k^i$ is independent, sampling $\bar{k}$ in \eqref{sam_k} is equivalent to sampling each element $k^i$ from
\begin{equation}
     p(k^i|s^i,\theta) \propto p(k^i|\theta) \ p(s^i|k^i,\theta) \ . 
\end{equation}
In both \eqref{sam_k} and \eqref{sam_th}, the most computationally expensive step is calculating the state probabilities 
conditioned on $k^i$ and $\theta$, $p(s^i|k^i,\theta)$, pseudocode for it is provided in Algorithm \ref{alg:J-MJP}. 

\begin{algorithm}[H]
\SetAlgoNoLine
\caption{State probabilities for J-MJP. Calculates the state probabilities, $p(s^i|k^i,\theta)$ as in \eqref{ME_uniform}.}\label{alg:J-MJP}
\KwIn{The dynamical parameters $\theta$, the initial probability vector $\bar{\rho}(0)$, the observed states $\bar{s} = \{s^1,s^2, ... s^I\}$, and the respective numbers of jumps  $\bar{k} = \{k^1,k^2, ... k^I\}$. }
\KwOut{An array, $\varphi$, of size $I$ (number of observed states) elements, whose $i$-th element is $p(s^i|k^i,\theta)$ .}
 \vspace{0.1cm}
 \hrule
 \vspace{0.1cm}
From $\theta$ calculate $\Omega_\theta$ and $\matth{B}$ \;
Set $\varphi$ as an empty array of $I$ elements\;
Set $k = 0$ and $\bar{\rho}^* = \bar{\rho}(0)$ \;
\While{$k \leq \max\limits_i{k^i}$}{
\ForEach{$i: k^i \compeq k$}{
$\varphi[i]  = \ p(s^i|k^i,\theta) =  (\bar{\rho}^*)_{s^i}$\;
    }
Set $k\pluseq 1$ \;
 Set $\bar{\rho}^* = \bar{\rho}^* \matth{B}$ \;
}
\Return $l$
\end{algorithm}

Rather than the complete set of $ \bar{\rho}(0)\matth{B}^k$ required by T-MJP, algorithm \ref{alg:J-MJP} only requires saving the vector $\bar{\rho} \matth{B}^k$, denoted in the pseudocode as $\bar{\rho}^*$. Its outer loop is run $\max\limits_i  k^i$ times. Given that $ k^i$ is Poisson distributed at rate $T^i \Omega_\theta$ \change{(see Sec. \ref{sec:forward})} and $\Omega_\theta$ scales with $\max\limits_{n}{|a_{nn}|}$, the number of outer loop calls scales with $O(T \max\limits_{n}{|a_{nn}|})$, with $T$ being the largest collection time $T = \max\limits_i T^i$. Meanwhile, the most computationally intensive operation within each loop iteration is multiplying $\bar{\rho}$ by $\matth{B}$, which scales in time as $O([\matth{B}]) = O([\matth{A}])$. Therefore the total time scaling of the state probability computation is $O(T \max\limits_{n}{|a_{nn}|} [\matth{A}])$. Under conditions a) and b) mentioned in the introduction, this is equivalent to $O(N^2)$.

\subsubsection{Method \#3 -- Remarginalized MJP method (R-MJP)}\label{sec:BayesR-MJP}

Eliminating the numerous samples of $\bar{k}$ from Method \#2, Method \#3 (R-MJP) directly approximates the full sum over $k$ in \eqref{ME_expansion}. That is, we obtain the probability vector at the observation time $\bar{\rho}(t)$, by truncating the sum over $k$ in \eqref{ME_expansion} at a cutoff $k^\ast$:
\begin{equation}\label{R-MJP_expansion}
\bar{\rho}(t) \approx \sum_{k=0}^{k^\ast} \bar{\rho}(0) \matth{B}^k \ \text{Poisson}(k|\Omega_\theta t) + \bar{\rho}(0) \matth{B}^k \  \varepsilon_{k^\ast}     \ ,
\end{equation}
where $ \varepsilon_{k^\ast} = 1 - \sum_0^{k^\ast} \ \text{Poisson}(k|\Omega_\theta t)  $, is the upper bound in the approximation error. 
\change{In} practice the goal is to calculate the probability vectors at the collection times $T^i$.
For practical reasons, we choose $k^\ast = \Omega_\theta t + 6 \sqrt{\Omega_\theta t}$, guaranteeing a modest error bound $ \varepsilon_{k^\ast} < 10^{-5}$ for $\Omega_\theta t >10$. While implementing the approximation in \eqref{R-MJP_expansion} is straightforward, we provide pseudocode detailing how to \change{efficiently} calculate the summation in \eqref{R-MJP_expansion} in Algorithm \ref{alg:R-MJP}.

\begin{algorithm}[H]
\SetAlgoNoLine
\caption{R-MJP for calculating the final probability vector $\bar{\rho}(t)$ as in \eqref{R-MJP_expansion}}\label{alg:R-MJP}
\KwIn{The dynamical parameters $\theta$, the initial probability vector $\bar{\rho}(0)$ and the time $t$.}
\KwOut{The final probability vector $\bar{\rho}(t)$ .}
 \vspace{0.1cm}
 \hrule
 \vspace{0.1cm}From $\theta$ calculate $\Omega_\theta$ and $\matth{B}$ \;
Set $k = 0$ and $P_\text{Poisson} = e^{- \Omega_\theta t}$ \;
Set $P_\text{Cumulative} = P_\text{Poisson} $ \;
Set $\bar{\rho} = \bar{\rho}(0)$ \;
Set $\bar{\rho}^* = P_\text{Poisson} \bar{\rho} $ \;

\While{$k \leq \Omega_\theta t + 6\sqrt{(\Omega_\theta t)}$}{
    Set $k \pluseq 1$ \;
    Set $\bar{\rho} = \bar{\rho} \matth{B}$ \;
    Set $P_\text{Poisson} *= \frac{\Omega_\theta t}{k}$ \;
    Set $P_\text{Cumulative} \pluseq P_\text{Poisson} $ \;
    Set $\bar{\rho}^* \pluseq P_\text{Poisson}*\bar{\rho}$ \;     
}
Set $\bar{\rho}^* \pluseq (1 - P_\text{Cumulative})*\bar{\rho}$ \;
\Return $\bar{\rho}^*$
\end{algorithm}

In order to calculate the probability vector at the observation time $\bar{\rho}(t)$ from $\bar{\rho}(0)$, the outer loop in Algorithm \ref{alg:R-MJP} is set to run $k^\ast = \Omega_\theta t + 6 \sqrt{\Omega_\theta t }$ times. Similarly to the previous method, this scales as $O(T \max_n |a_{nn}|)$. Within this loop, the most computationally expensive task is multiplication of an array of $N$ elements by $\matth{B}$, with scales with $O([\matth{B}])$. Consequently, the overall time cost scaling is $O(\max_n |a_{nn}|[\matth{B}])$. Again, when conditions a) and b) in Sec. \ref{sec:Bayes} are met, this overall cost becomes $O(N^2)$, while the Algorithm \ref{alg:R-MJP} requires only saving a single array of size $N$ besides the matrix $\matth{B}$, giving memory scaling $O(N)$. 

Note that unlike the previous methods, which require latent variable sampling, this method allows us to directly approximate the solution of \eqref{differential_ME} while avoiding the matrix exponential in \eqref{ME_solution}. Notably, that the number of virtual jumps, $k$, follows a Poisson distribution with rate $\Omega_\theta T$, which we can interpret $1/\Omega_\theta$ as something like an adaptive time grid. Subsequent methods which directly approximate \eqref{differential_ME} differ from R-MJP by explicitly segmenting time into steps, $\Delta t$, which must be chosen by the user. By comparison, Method \#3 adaptively determines $k^\ast$ based the system's dynamics and specifies an error upper bound, $\varepsilon_{k^\ast}$, with no additional computation.

\subsubsection{Method \#4 -- Runge-Kutta}\label{sec:RK}

The most conceptually straightforward technique to avoid the matrix exponential in \eqref{ME_solution} is to numerically integrate the trajectory-marginalized equation \eqref{differential_ME}. Runge-Kutta, designed for ordinary differential equations \cite{Butcher00}, balances stability and accuracy flexibly based on a chosen order of expansion and time step $\Delta t$. 
For example, first order Runge-Kutta reads % \eqref{differential_ME} 
\begin{equation}\label{Eulerstep}
    \bar{\rho}(t+\Delta t) \approx \bar{\rho}(t)  +  \Delta t \dv{\bar{\rho}}{t} =  \bar{\rho}(t)  +  \Delta t \ \bar{\rho}(t) \matth{A} \change{(t)} \ .
\end{equation}
Thus, we evolve the initial vector, $\bar{\rho}(0)$, and compute successive points along the solution curve until a final time $T$, {\it i.e.}, compute $\{\bar{\rho}(\Delta t),\bar{\rho}(2\Delta t),\bar{\rho}(3\Delta t), \ldots, \bar{\rho}(T)\}$ iteratively. This is equivalent to first order Taylor on the exponential at each time step. 

Avoiding numerical inaccuracy and instability often warrants higher-order methods. A common choice is the fourth-order Runge-Kutta method (RK4) \cite{Butcher00}.   

In RK4, we write
\eqref{differential_ME} as
\begin{equation}\label{RKstep}
    \bar{\rho}(t+\Delta t) \approx \bar{\rho}(t)  + \frac{1}{6}(\delta_1 + 2\delta_2 + 2\delta_3 + \delta_4) \ ,
\end{equation}
where
\begin{equation}\label{RKintermediate}
    \begin{split}
\delta_1 &= \Delta t \ \bar{\rho}(t) \matth{A}\change{(t)} \ ,  \\
\delta_2 &= \Delta t \ \left( \bar{\rho}(t) + \frac{\delta_1}{2} \right) \matth{A}\change{(t+\nfrac{\Delta t}{2})} \ , \\
\delta_3 &= \Delta t \  \left( \bar{\rho}(t) + \frac{\delta_2}{2} \right) \matth{A}\change{(t+\nfrac{\Delta t}{2})}  \ , \\
\delta_4 &= \Delta t \  \left( \bar{\rho}(t) + \delta_3 \right) \matth{A}\change{(t+{\Delta t})}     \ .
    \end{split}
\end{equation}
and iteratively evolve from $t=0$ towards $t = T$. This reduces the error to $\Delta t^4$ \change{(see Ref.\cite{Butcher00})} at each step. 
\change{We made the time dependence in \eqref{RKintermediate} explicit (writing $\matth{A}(t)$) because this method can be used for transition rates that change in time, in opposition to the previous methods relying on the expansion in \eqref{ME_expansion} that assumes $\matth{A}$ is constant in time.}

When implementing RK4, numerical stability considerations dictate that $\Delta t \leq \frac{1}{\max\limits_{n}{|a_{nn}|}} $. 
Thus, the number of iterations required in computing the likelihood is given by $T \times \max\limits_{n}{|a_{nn}|}$. Since evolving by $\Delta t$ requires 4 matrix-vector products \eqref{RKintermediate}, this scales in time as $O([\matth{A}] )$.
{When conditions a) and b) mentioned Sec. \ref{sec:Bayes} are met, it follows that the Runge-Kutta solution takes a time of order $O(N^2)$. Besides storing $\matth{A}$, no more than 5 arrays of size $N$ need to be saved at a time --- $\delta_1$ to $\delta_4$ in \eqref{RKintermediate} and $\bar{\rho}(t)$. Thus leading to the value found in Table \ref{tab:requirements}. }

\subsubsection{Method \#5 -- Krylov subspace}\label{sec:Krylov}

The Krylov subspace technique \cite{Vo17,gaudreault18,Vo19} evolves the initial probability vector $\bar{\rho}(t)$ through time by approximating the matrix exponential over a short time step $\Delta t$, $\bar{\rho}(t+\Delta t) = \bar{\rho}(t) e^{\matth{A}\Delta t}$. In keeping with the other methods, it still avoids the full matrix exponential $e^{\matth{A} \Delta t}$. \change{It does so} by constructing a Krylov subspace at each time step, $\mathcal{K}_\kappa(\matth{A},\bar{\rho}(t))$, spanned by $\{ \bar{\rho}(t), \bar{\rho}(t) \matth{A}, \bar{\rho}(t) \matth{A}^2, $ $ \bar{\rho}(t) \matth{A}^3, \ldots, \bar{\rho}(t) \matth{A}^{\kappa-1} \}$, where the size of the subspace, $\kappa$, is a predefined integer which serves a role similar to the approximation order in Runge-Kutta. \change{Note that, as with Methods \#1 -- \#3, the Krylov subspace requires  that $\matth{A}$ is constant in the interval $(t,t+\Delta t)$. }
 
Although the vectors spanning the subspace are clearly defined, we require an orthonormal basis \( \{\bar{q}_1, \bar{q}_2, \ldots, \bar{q}_\kappa\} \) for \( \mathcal{K}_\kappa(\matth{A},\bar{\rho}(t)) \), in order to project \( \matth{A} \) into the subspace. This projection is given by 
\begin{equation}\label{Krylov_projection}
    \mat{H}^\dag = \mat{Q}^\dag \matth{A} \mat{Q} \ , 
\end{equation}
where $\mat{Q}^\dag$ represents the transpose of $\mat{Q}$ and $ \mat{Q} $'s $i$-th column is $\bar{q}_i^\dag$. Both $\mat{H}$, along with the basis  \( \{\bar{q}_1, \bar{q}_2, \ldots, \bar{q}_\kappa\} \), are obtained using the Arnoldi algorithm \cite{Arnoldi51}. To make its computational cost clear, we describe the Arnoldi algorithm in Algorithm \ref{alg:ArnoldiKrylov}.

\begin{algorithm}[H]
\SetAlgoNoLine
\caption{Arnoldi algorithm for obtaining the orthonormal basis and projection of the generator matrix onto the Krylov subspace.}\label{alg:ArnoldiKrylov}
\KwIn{The generator matrix $\matth{A}$, the initial probability vector $\bar{\rho}(t)$ for the interval $[t,t+\Delta t]$, and the Krylov subspace size $\kappa$.}
\KwOut{$\mat{Q}$, whose columns are an orthonormal basis of $\mathcal{K}_\kappa(\matth{A},\bar{\rho}(t))$ and $\mat{H}$, the projection of $\matth{A}$ into the Krylov subspace~\eqref{Krylov_projection}.}
 \vspace{0.1cm}
 \hrule
 \vspace{0.1cm}
 From $\bar{\rho}(t)$ obtain $\bar{q}_1 =  \bar{\rho}(t)/ || \bar{\rho}(t)||_2$, with $ ||\cdot||_2$ representing the 2-norm \;
 Set $\mat{H}$ as a $\kappa \times \kappa$ of elements $h_{ij}$ all equal to zero \;
 \ForEach{$i: 2, 3, \ldots, \kappa $}{
   Set $\bar{q}_i = \bar{q}_{i-1} \matth{A}$  \;
    \ForEach{$j: 1, \ldots, i-1 $}{
      Set $h_{j(i-1)} = \langle \bar{q}_j, \bar{q}_i \rangle$, with $\langle \cdot,\cdot \rangle$ denoting the inner product \;
       Set $\bar{q}_i = \bar{q}_i - h_{j(i-1)}\bar{q}_j$ \;
    }
  Set $h_{i(i-1)} = ||\bar{q}_i||_2$ \;
  Set $\bar{q}_i = \frac{\bar{q}_i}{h_{i(i-1)}}$ \;
 }
 \Return $\mat{Q}$, a matrix whose $i$-th column is $\bar{q}_i^\dag$, and $\mat{H}$ 
\end{algorithm}

From \eqref{Krylov_projection}, we can approximate the matrix exponential in \eqref{ME_solution} as
\begin{equation}
    \bar{\rho}(t+\Delta t) = \bar{\rho}(t) e^{\matth{A} \Delta t} \approx  \bar{\rho}(t)  \mat{Q} (e^{\mat{H} \Delta t})^\dag \mat{Q}^\dag \ . 
\end{equation}
Now since the columns of $\mat{Q}$ form an orthonormal basis, $\mat{Q}^\dag = \mat{Q}^{-1}$. Using also the fact that the first basis vector is parallel to $\bar{\rho}(t)$, we can simplify further 
\begin{equation}\label{Krylov_expansion}
    \bar{\rho}(t+\Delta t) = \bar{\rho}(t) e^{\matth{A} \Delta t} \approx  ||\bar{\rho}(t)||_2  \left( (\mat{Q} e^{\mat{H} \Delta t})_{[:,1]}  \right)^\dag\ , 
\end{equation}
with the subscript $[:,1]$ meaning the first column of the matrix. 

Naturally, the choice of $\kappa$ and $\Delta t$ contribute to the approximation error and computational cost of Method \#5. In particular Krylov allows larger $\Delta t$ than RK4  --- we have found that  $\Delta t = \frac{2\kappa}{\max\limits_{n}{|a_{nn}|}} $ leads to a stable integrator. The number of iterations required to compute the $\bar{\rho}(T)$ \eqref{ME_solution} by sequentially approximating \eqref{Krylov_expansion} is then given by $\frac{T}{2 \kappa} \times \max\limits_{n}{|a_{nn}|}$. 
However, larger values of $\kappa$ also lead to larger matrices $\mat{Q}$ and $\mat{H}$, leading back to the original problems with the matrix exponential. For our later benchmark, we fix $\kappa = 20$.

There are three calculations at each time step in \eqref{Krylov_expansion}. First compute $\mat{Q}$ and $\mat{H}$ through Algorithm \ref{alg:ArnoldiKrylov}. Each iteration requires $\kappa$ matrix times vector products, scaling as $O(\kappa [\matth{A}])$, and $\frac{\kappa (\kappa-1)}{2}$ dot products, $O(\kappa^2 N)$. Next, calculate $e^{H \Delta t}$, which scales as $O(\kappa^3)$. Finally, perform the matrix multiplication in \eqref{Krylov_expansion}, which scales as $O(\kappa N)$. Thus the total time scaling becomes $O\left(T  {\max\limits_{n}{|a_{nn}|}} ([\matth{A}] + \kappa N +\kappa^2) \right)$. 

At each iteration three objects are stored: $\matth{A}$, $\mat{Q}$, and $\mat{H}$. Therefore the memory cost scales as $O(\kappa^2 + \kappa N + [\matth{A}])$. 
Again assuming a) and b) mentioned in the Sec. \ref{sec:Bayes} are satisfied, the total time and the memory requirements scale as $O(\kappa^2+\kappa N^2)$ and $O(\kappa N)$, respectively. These together determine the scaling per iteration presented in Table \ref{tab:requirements}.

Note that we usually use a fixed $\kappa$ and observe the scaling for $N$, as such we are most interested in $\kappa \ll N$, but we will not take that limit in scaling explicitly as of now. This happens because, in practical terms, the presence of $\kappa$ leads to overhead in the total time calculation that will be relevant in our benchmarking. However one could expect that when such limit is reached the scaling would become $O(N^2)$ for computational time and $O(N)$ for memory, similarly to the previous methods.

\section{Results}\label{sec:results}

With all of the methods described, we now compare them on three systems of interest we present below (Sec. \ref{sec:examples}). Benchmarking of the different methods is given \change{in} Sec. \ref{sec:benchmarking}.

\subsection{Examples}\label{sec:examples}

\begin{figure}
\begin{center}
   \includegraphics[width=.8\textwidth]{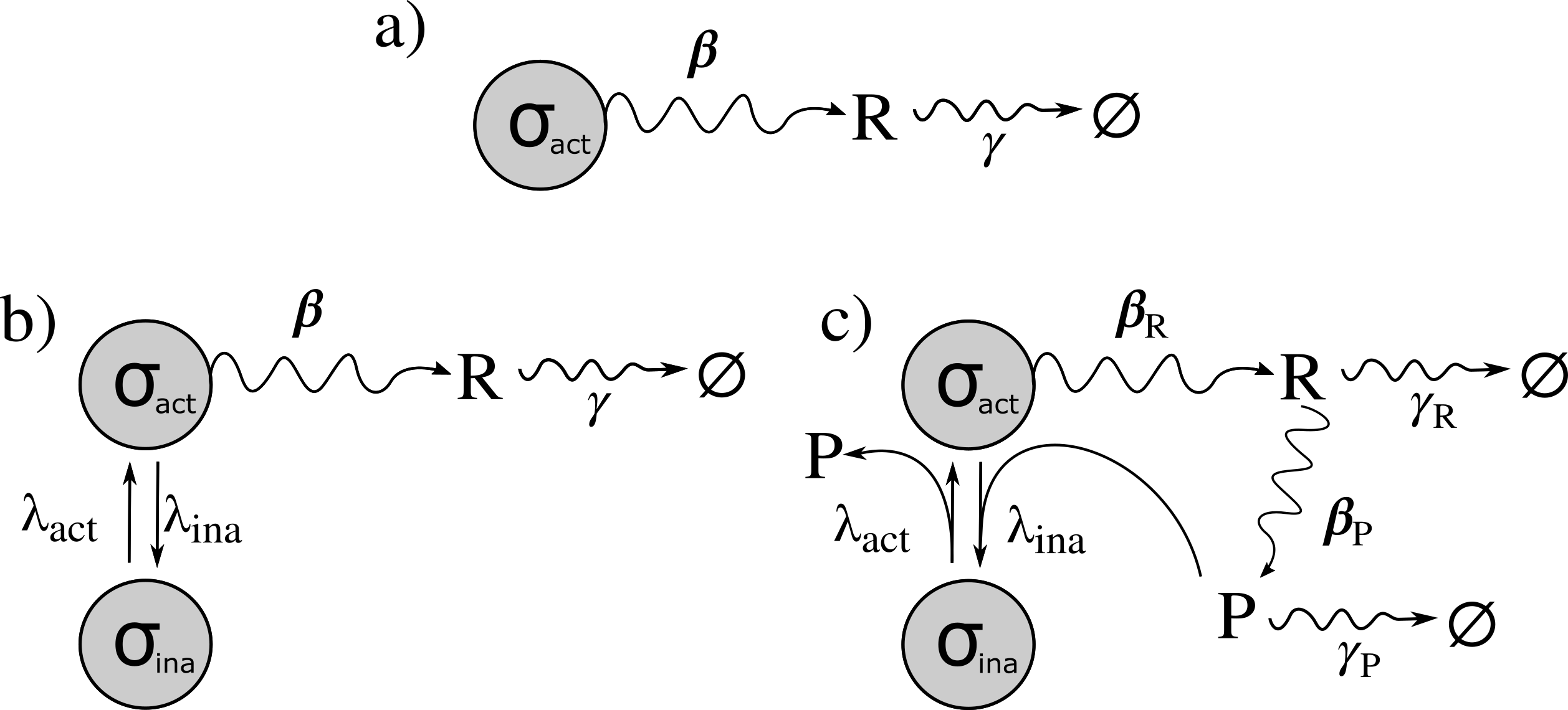}
\end{center}
\caption{Cartoon representation of dynamical systems described in Sec. \ref{sec:examples}. Panel {\bf a)} represents a birth-death process of $R$ as described in Sec. \ref{sec:Birth-Death}, panel {\bf b)} represents a system that switches between two states, only one of them, \change{the active one} $\sigma_\text{act}$, produces $R$, as in Sec. \ref{sec:Gene_example}, and panel {\bf c)} represents an autoregulatory system, with one state, $\sigma_\text{act}$, generating a product $R$ which generates a second product $P$, which subsequently plays a role in inactivating the system by moving it to a state, $\sigma_\text{ina}$,  where $R$ is not produced, as in Sec. \ref{sec:autorel}.
}\label{fig:cartoon}
\end{figure}

As a basis for our benchmark, we describe some examples of dynamics mediated by \eqref{differential_ME}, inspired by selected literature. An illustrative cartoon of each system is found in Fig. \ref{fig:cartoon}.  

In the subsections dedicated to each example below, we will: first motivate the model and describe the required parameters, $\theta$; second, build the state space and formulate the transition rate matrix, $\mat{\Lambda}$; third, we describe the simulation of synthetic experiments producing the data to be used in our benchmark in Sec. \ref{sec:benchmarking}.

\subsubsection{Birth-death process}\label{sec:Birth-Death}
The birth-death process is relevant across population dynamics. It models stochastic production (or birth) of a species $R$ at constant rate $\beta$ that can be degraded (or die) at a rate $\gamma$. Thus the full set of parameters is simply $\theta = \left\{ \beta,\gamma \right\}$. A representation of this process is depicted in Fig. \ref{fig:cartoon} {\bf a)}. Expressed in the language of chemical reactions we have 
\begin{equation}\label{BD_chem}
\begin{split}
    \emptyset & \xrightarrow{\beta} R \\
    R & \xrightarrow{\gamma} \emptyset \ .
\end{split}
\end{equation} 
\change{For example} in molecular biology, we can model an actively transcribing DNA sequence producing one RNA \cite{Kilic23} with rate $\beta$ degraded at rate $\gamma$. Data for inference on this system would be the population of $R$ at different time points across a number of identical DNA sequences.

We write the elements of the state space $\sigma_n$ to represent an $R$ population of $n$, with $N$ being a upper cut-off in the $R$ number. 
While selection of the upper cut-off can be worthy of independent investigation --- see \emph{e.g.} \cite{Munsky06,Vo19,munsky2018distribution} --- here, and in all following examples, we determine $N$ in a data-driven manner. To set the cut-off, we assume that the probability for $R$ counts larger than double the highest count found in the data is effectively zero. 
In this case, the elements $\lambda_{nm}(\theta)$ of the  transition rate matrix $\mat{\Lambda}$ are given by
\begin{equation}\label{BD_def}
    \lambda_{nm}(\theta) \doteq 
    \begin{cases}
    \beta \quad & \text{if} \quad  m=n+1 \\ 
    \gamma n  \quad & \text{if} \quad m= n-1 \\
    0 \quad \quad & \text{otherwise}
    \end{cases} \ .
\end{equation}
We note that $\mat{\Lambda}$ is sparse by construction, and each row has, at most, two non-vanishing elements. Since $\max\limits_n a_{nn}$ was an important quantity when discussing the methods' time scaling, it is straightforward to see from \eqref{A_def} that, $\max\limits_n |a_{nn}| = \max\limits_n( \beta+\gamma n) = \beta+\gamma (N-1)$, which scales as $O(N)$.

In our synthetic experiments for this process, we fix the time units so the ground truth degradation rate is one $\gamma = 1$ or, equivalently, time has units of $1/\gamma$. 
Crucially, this is merely a choice of parameters for the synthetic experiment and we do not assume any knowledge of $\gamma$ \emph{a priori}. 
All simulations start with zero population of $R$, and we observe the final state at various time points $T^i$, evenly spaced from $0.5$ to $5$. For each of these time points, we conduct 300 individual simulations, yielding a total dataset of $I=3000$ observations. 
We perform synthetic experiments for each value of the production rate $\beta$ from the set $\{ 50,100, 200,400, 800 \}$, selected because the population at equilibrium fluctuates around $\beta/\gamma$ allowing us to assess the scaling \change{(on time and memory)} of each algorithm with the number of states \emph{necessary} for accurate inference, spanning a range from hundreds to thousands of states.

\subsubsection{Two state birth-death}\label{sec:Gene_example}

Next we explore a two state birth-death reaction network as a simple extension of the birth-death process where the system's production stochastically deactivates. We reuse $\beta$ for the production rate in the system's active state, $\sigma_\text{act}$. The product $R$ degrades at a constant rate $\gamma$ just as before. Additionally, the system deactivates --- transitions from $\sigma_\text{act}$ to an inactive state $\sigma_\text{ina}$ --- at a rate $\lambda_\text{ina}$ and reactivates at a rate $\lambda_\text{act}$. A representation of this system is depicted in Fig. \ref{fig:cartoon} {\bf{b)}}.

The full set of kinetic parameters is denoted by $\theta = \{\beta,\gamma, \lambda_\text{act}, \lambda_\text{ina} \}$, and the system's state is described by two variables: the population of $R$, denoted as $n_R$, and the gene state, $\sigma_G$, which can be either active, $\sigma_G = 0$ representing $\sigma_\text{act}$, or inactive, $\sigma_G=1$ representing $\sigma_\text{ina}$.

This is equivalent to so-called `bursty' expression of a gene with one active --- meaning, RNA producing --- and one inactive state. Expressed in the language of chemical reactions we write
\begin{equation}\label{2S_chem}
\begin{split}
    \sigma_\text{act} & \xrightarrow{\beta} R + \sigma_\text{act}\\
    R & \xrightarrow{\gamma} \emptyset \\ 
    \sigma_\text{act} & \xleftrightharpoons[\lambda_\text{act}]{\lambda_\text{ina}} \sigma_\text{ina} \ .
\end{split}
\end{equation}

Since \change{we must choose a linear ordering of states, in order to write their probabilities as vector,} we ought to write our state space with a single index $n$. \change{We} index the states using $n = n_R + \sigma_G N_R$ with $N_R$  the upper cut-off for the $R$ population, and $N=2 N_R$. Similarly, we can recover the population of $R$, $n_R$, and the gene state $\sigma_G$ \change{from $n$} as $(n_R,\sigma_G) = \left(n \mod {N_{R}}, \left\lfloor \frac{n}{N_R} \right\rfloor \right)$, where $x \mod q$ represents the remainder obtained when $x$ is divided by $q$ and $\lfloor x \rfloor $ represents the largest integer smaller than $x$.
Here, the transition matrix elements are given by
\begin{equation}\label{Gene_def}
    \lambda_{nm}(\theta) \doteq 
    \begin{cases}
    \beta \quad & \text{if} \quad  m = n+1 \ \text{and} \ n<{N_{R}} \\ 
    \gamma \ n_{R} \quad & \text{if} \quad m=n-1 \\
    \lambda_\text{act}   \quad & \text{if} \quad m=n+{N_{R}} \\
    \lambda_\text{ina}   \quad & \text{if} \quad m=n-{N_{R}} \\
    0 \quad \quad & \text{otherwise}
    \end{cases} \ .
\end{equation} 
Since all elements $\lambda_{nm}$ are positive and going back to the definition of $a_{nm}$ in \eqref{A_def} we can see that  $\max\limits_n |a_{nn}| \leq \max\limits_n( \beta+\gamma n_R + \lambda_\text{act} + \lambda_\text{ina}) = \beta+\gamma (N_{R}-1) + \lambda_\text{act} + \lambda_\text{ina}$, scaling as $O(N)$.

Just as we did for the previous system, we set $\gamma =1$ in our synthetic experiment thereby setting the units.
In this cases, the production rate $\beta$ takes the values in $\{ 100, 200, 400, 500 \}$ enabling us to \change{observe how the time and memory scales within each} of the methods with the number of states, since the steady-state amount of product for the previous example, $\beta/\gamma$, is observed in the limiting case where the system never moves from the active state (\emph{i.e.}, if $\lambda_\text{ina} = 0$). The activation and deactivation rates, $\lambda_\text{act}$ and $\lambda_\text{ina}$, are chosen to be on the same order as the measurement time, $\lambda_\text{act} = 2$ and $\lambda_\text{ina}=1$. The observation points, $T^i$, are evenly spaced from $0.5$ to $10$, leading to a total of $I=6000$ observations. 
The observations (data) are the population of $R$, $n_R$, at the different time end points. However unlike in the previous example there is no direct observation of the $\sigma_\text{act}$ or $\sigma_\text{ina}$ state. Hence, when calculating the likelihood of observation one has to sum the probabilities of states $(n_R,0)$ and $(n_R,1)$.

\subsubsection{Autoregulatory gene network}\label{sec:autorel}

Another scenario worth studying is the model for an autoregulatory gene network. In such a system, RNA is produced that is later translated into a protein which, in turn, suppresses its own production --- see Fig. \ref{fig:cartoon} {\bf c)}. As a result, the system is capable of maintaining a controlled level of both transcribed RNA and synthesized protein within cells \cite{Sukys22}.

In order to model this system, we consider an active state, $\sigma_\text{act}$, and an inactive state, $\sigma_\text{ina}$. In the active state, the system produces a component $R$ at a rate of $\beta_R$. Meanwhile, each copy of $R$ can produce another component $P$ at rate $\beta_P$ . This additional component $P$, in turn, inhibits the production of the first product $R$. 

Expressed in the language of chemical reactions we have
\begin{equation}\label{STS_chem}
\begin{split}
    \sigma_\text{act} & \xrightarrow{\beta_R} R + \sigma_\text{act} \\
    R & \xrightarrow{\beta_P} R +P \\ 
    \sigma_\text{act} + P & \xleftrightharpoons[\lambda_\text{act}]{\lambda_\text{ina}} \sigma_\text{ina} \\
    R & \xrightarrow{\gamma_R} \emptyset \\
    P & \xrightarrow{\gamma_P} \emptyset \ .
\end{split}
\end{equation}

Thus, the set of kinetic parameters is denoted by $\theta = \{\beta_R, \beta_P,\gamma_R,\gamma_P, \lambda_\text{act} ,\lambda_\text{ina} \}$, and the system's state is described by three variables: the population of $R$, $n_R$; the population of $P$, $n_P$; and the gene state, $\sigma_G$. 
Similarly to how it was done in the previous example, we index the state space using $n = n_R + n_P N_R + \sigma_G N_R N_P$ which is inverted as $(n_R,n_P,\sigma_G) = \left(n \mod {N_R}, \left\lfloor\frac{n}{N_R} \right\rfloor \mod N_P, \left\lfloor\frac{n}{N_R N_P} \right\rfloor  \right)$, with $N_R$ and $N_P$ being the cutoff on the population of $R$ and $P$, respectively. Thus $N = N_R N_P$. 
The transition matrix elements are given by 
\begin{equation}\label{autoReg_def}
    \lambda_{nm}(\theta) \doteq 
    \begin{cases}
    \beta_R                \quad & \text{if} \ m = n+1, \ \left\lfloor\frac{n}{N_R} \right\rfloor \mod N_P \neq N_P-1, \ \text{and} \ n< N_R N_P \\ 
    \beta_P \ n_R              \quad & \text{if} \ m = n+N_R, \ (n \mod {N_R}) \neq N_R-1, \ \text{and} \ n< N_R N_P \\ 
    \gamma \ n_R               \quad & \text{if} \ m = n-1   \\
    \gamma \ n_P               \quad & \text{if} \ m = n-N_R\\
    \lambda_\text{act}     \quad & \text{if} \ m = n-{N_R N_P}+N_R \\
    \lambda_\text{ina} \ n_P   \quad & \text{if} \ m = n+{N_R N_P}-N_R \ \text{and}  \left\lfloor\frac{n}{N_R} \right\rfloor \mod N_P \neq N_P-1 \\
    0 \quad                \quad & \text{otherwise}
    \end{cases} \ .
\end{equation}
Here calculating the largest element of the transition rate matrix is not as trivial as in the previous examples, but it can be done by plugging the elements $\lambda_{nm}$ into \eqref{A_def} obtaining
\begin{equation}
    \max\limits_n {|a_{nn}|} \leq \beta_R + (\beta_P + \gamma) (N_R-1) + (\gamma +\lambda_\text{ina}) (N_P-1) + \lambda_\text{act} \ .
\end{equation}
Although not as straightforward as in the previous examples, this system scales better than linearly with the total size of the state space $N$. Later this will mean that the scaling in time will be considerably better than expected from Table \ref{tab:requirements}.

Here we use the degradation rate of $R$ to set our time units, $\gamma_{R} =1$. We explore the production rates for both $R$ and $P$ as $(\beta_R,\beta_P) \in \{(2.5,.125),(5,.25), (10,.5),(20,1)\}$. 
The activation and deactivation rates are set at $\lambda_\text{act} = .1$ and $\lambda_\text{ina}=0.05$ and the degradation rate of $P$ is given as $\gamma_P =.1$. 
Smaller production and degradation rates are chosen for $P$ to represent the greater stability and higher production cost of proteins over RNA.
Similarly to the previous example, the collection times are taken evenly spaced from $0.5$ to $10$, leading to a total of $I=6000$ observations. The data available are now the population of both $R$ and $P$, $n_R$ and $n_P$ respectively, at the final state. Similarly to the previous example. there is no direct observation of $\sigma_\text{act}$ or $\sigma_\text{ina}$ state. Hence, when calculating the likelihood of an observation one has to sum the probabilities of states $(n_R,n_P,0)$ and $(n_R,n_P,1)$.

\subsection{Benchmarking}\label{sec:benchmarking} 

To evaluate the efficiency of the Methods (\#2--\#5) in avoiding the matrix exponential for a variety of physically motivated generator matrices, we embed the matrix exponential as the key computational step in a Bayesian inference scheme. To be precise, we generate synthetic data $\bar{s}$ and perform inference by {sampling kinetic parameters, $\theta$ from the} posterior 
$p(\theta|\bar{s})$ with MCMC. The sampling scheme (see Supplemental Information \ref{SIsec: sampling} for details) is of little relevance to the matrix exponential, but we've made the source code publicly available on our GitHub \cite{github}.

Benchmarking each method serves two primary objectives: first, it aims to evaluate time scaling with total number of states for the benchmarked methods to compare them with the conventional matrix exponential \cite{Scipy_documentation}. Since the conventional matrix exponential is dramatically more costly than other methods, even for this simple system, we only include it in our benchmark for the birth-death process. 
These results are presented and discussed in Sec. \ref{sec:time}.

Second, to ensure that each method results in an equivalent modeling scheme, we present a comparison between the probability distributions over $\theta$ obtained through various methods in the Supplemental Information \ref{SIsec:cons}. Notably, even though the time needed to generate the samples which make up this distribution with the J-MJP (Method \#2) is not considerably worse than the other methods, it generates highly correlated samples, a problem whose implications we expand upon in Sec. \ref{sec:J-MJPac}.

\subsubsection{Time \change{scaling} comparison}\label{sec:time}

\begin{figure}%[h]
\begin{center}
   \includegraphics[width=.6\textwidth]{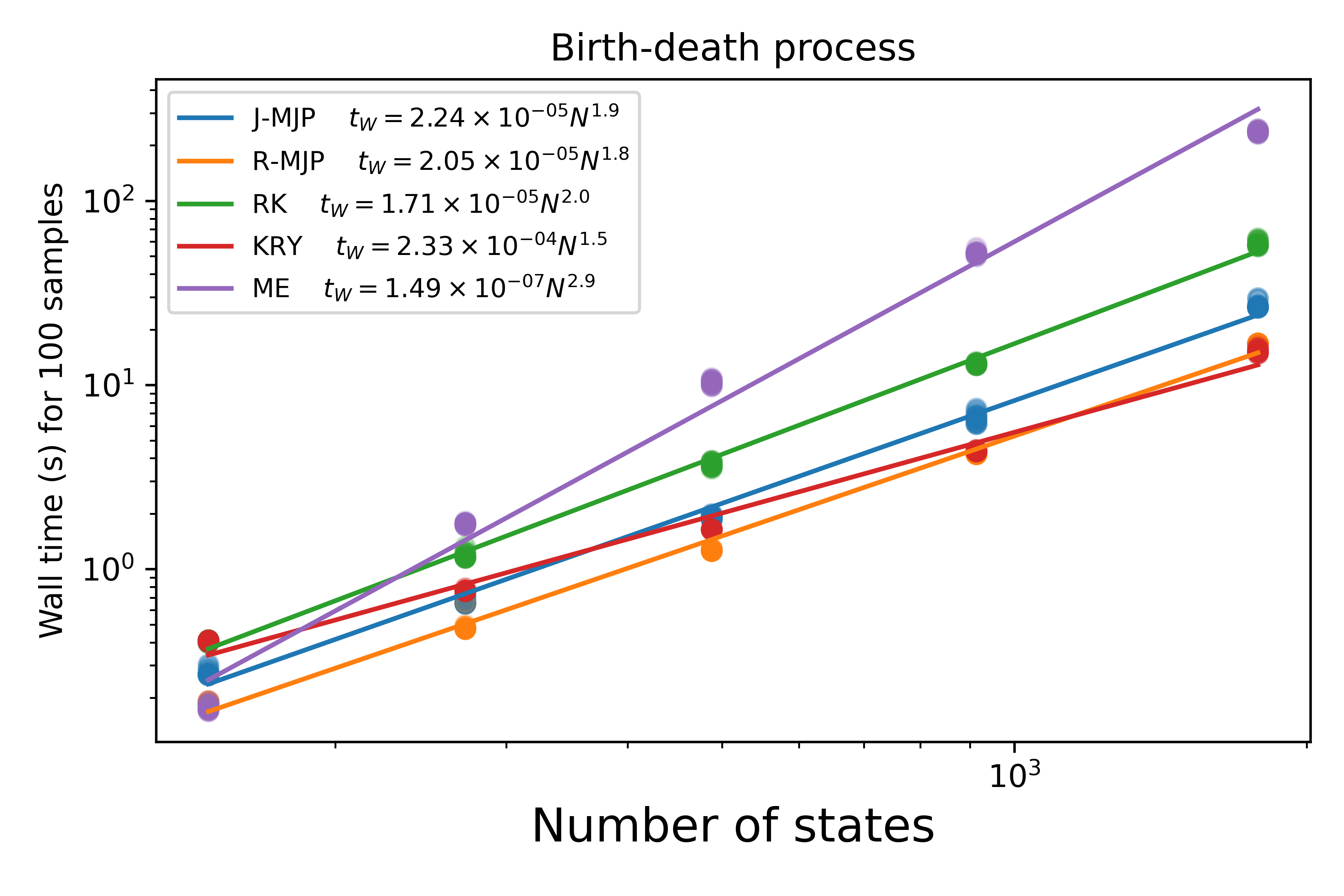} \vspace{-.25cm}
   
   \includegraphics[width=.6\textwidth]{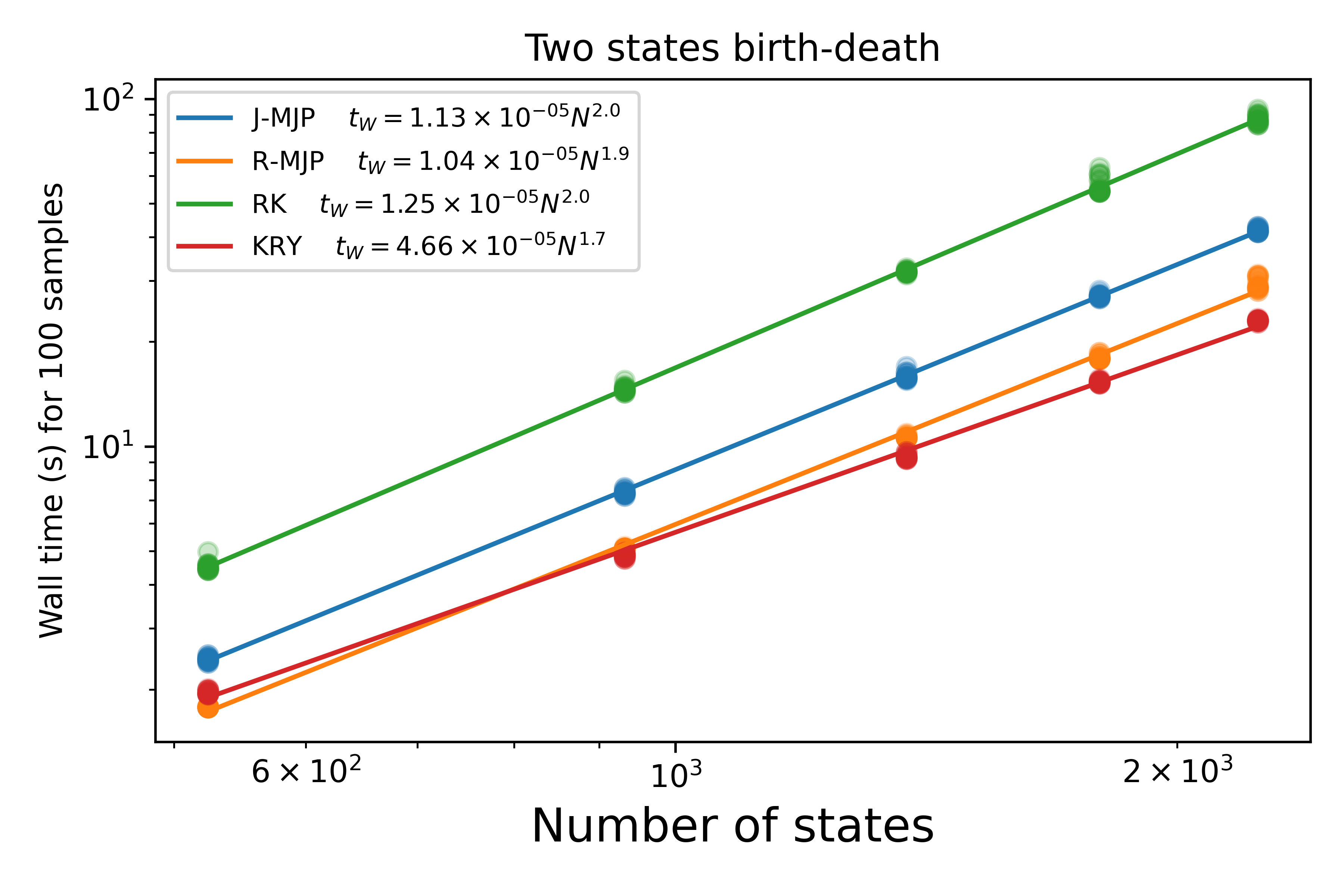}\vspace{-.25cm} 
   
   \includegraphics[width=.6\textwidth]{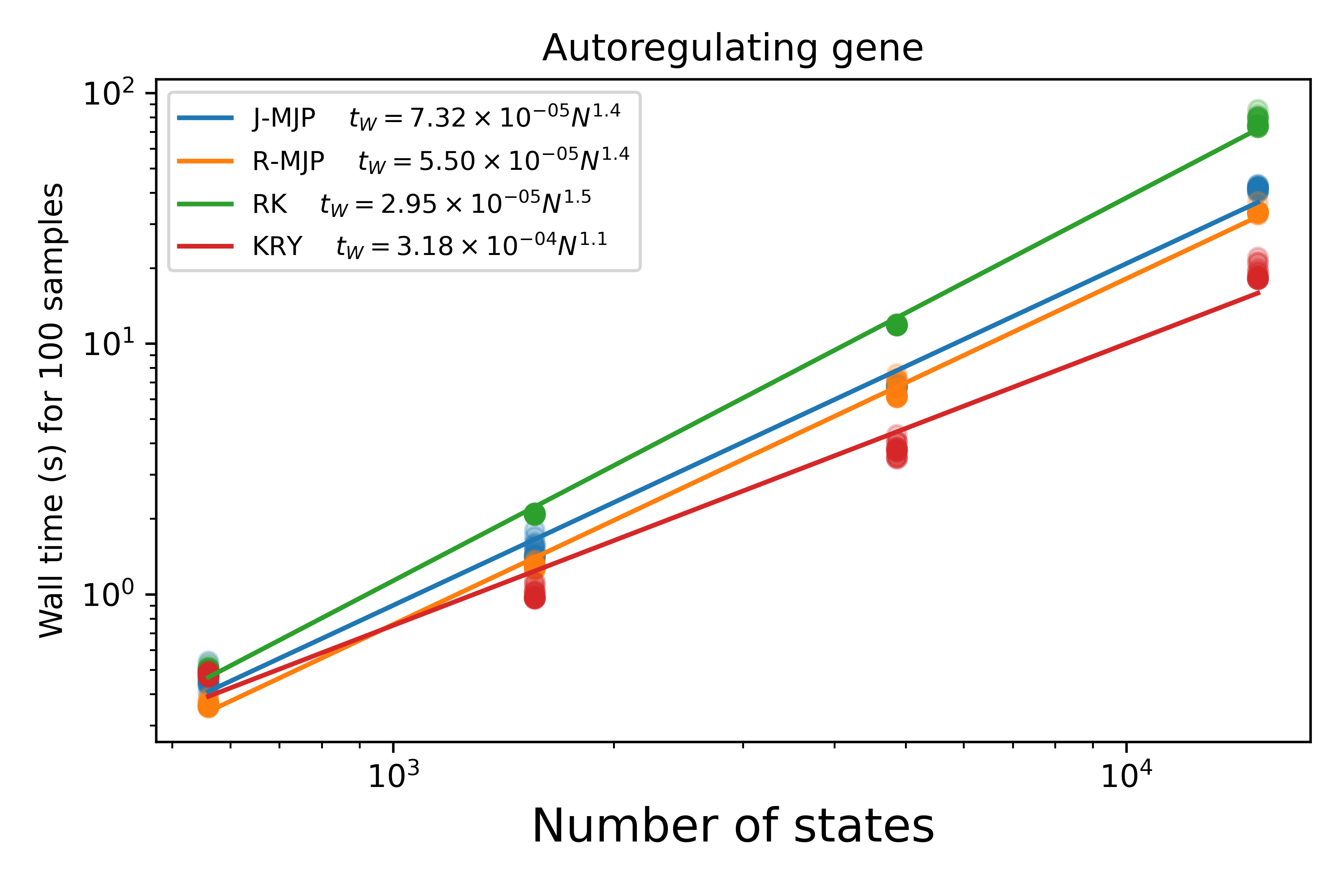}
\end{center}\vspace{-1cm}
\caption{ We recorded the wall time, $t_W$, necessary to obtain 100 samples from the posterior in relation to the total number of states. 
A comparison of J-MJP, R-MJP, RK4, and Krylov \change{(KRY)} for all three examples is presented above. In the birth-death process (Sec. \ref{sec:Birth-Death}) is presented on the \change{top} panel, the two states birth-death (Sec. \ref{sec:Gene_example}) is presented in the middle panel example and for the autoregulatory gene model (Sec. \ref{sec:autorel}) is presented in the \change{bottom} panel. In the birth-death example, we also compare to the conventional implementation of the matrix exponential (ME). 
When considering overall performance, the contest narrows down between R-MJP and Krylov, with Krylov tends to exhibit superior performance, especially as the state spaces increase in size.
}\label{fig:scaling}
\end{figure}

To compare the time needed to calculate the final distribution (thus avoiding the matrix exponential in \eqref{ME_solution}) used for each method, we measure the wall time required to sample 100 values for all kinetic parameters from across datasets (each demanding different total state space size). To obtain some statistics, we repeat this procedure 20 times. The results can be seen in Fig. \ref{fig:scaling}. We focused \change{on comparing the time scaling between methods.} conducting tests under conditions where MCMC samplers had already converged to the posterior by initializing it at the simulation's ground truth. 

Our Python code, available in our repository \cite{github}, leverages the Numba compilation tool \cite{Numba} and a package enabling sparse matrix manipulation in Numba-compiled code \cite{smn}.
The benchmark was conducted on a system equipped with an Intel(R) Core(TM) i7-7700K CPU operating at a base frequency of 4.20 GHz. The system is outfitted with 16 GB of RAM.

Fig. \ref{fig:scaling} clearly illustrates the expected quadratic scaling of RK4, J-MJP, and R-MJP relative to the total number of states 
\change{in each of the three examples under study}.  In these cases, we observe a slightly smaller exponent due to necessary overhead.   
Nevertheless, this overhead is anticipated to diminish in significance with an increase in state space size. The Krylov scaling is somewhat less straightforward as the method has a more complex dependence on the number of states, as discussed in Sec. \ref{sec:Krylov} and Table \ref{tab:requirements}, as such needs a considerably larger number of states to actually reach a regime of quadratic scaling. 

When studying the autoregulatory gene system, the exponents of the power law relating the wall time and number of states are smaller due to the less straightforward relationship between the largest element of $\matth{A}$ and the total number of states, as mentioned in Sec \ref{sec:autorel}. 
In the same figure we also show, for the birth-death process Sec. \ref{sec:Birth-Death}, how all of these methods dramatically outperform regular matrix exponential implementation.

\change{The comparison between the time scaling between Methods \#2 -- \#5 and the regular matrix exponential} was not implemented in the two other systems, as the difference is even more dramatic.  The competition for overall best method comes between R-MJP and Krylov, with the tendency of Krylov to better perform at larger state spaces. 
\change{
This difference in performance can be attributed to the fact that, in smaller state spaces, the time required for allocating memory to store the Krylov subspace (as detailed in Algorithm \ref{alg:ArnoldiKrylov}) and for exponentiating the $\kappa \times \kappa$ projection matrix, $\mathbf{H}$, is significant compared to the time needed for the vector-matrix product computation. However, as the state space expands, the relative time consumed for memory allocation and matrix exponentiation diminishes, making the Krylov method increasingly more efficient. Nevertheless, despite the size of the state space, Krylov still \emph{requires} arbitrary choices of the subspace size, $\kappa$ and the time step (with the latter being also required in RK4). In contrast, R-MJP (as well as T-MJP and J-MJP) have an the effective time step, $1/\Omega_\theta$, endowed by the system.
}

\change{
Determining the state space size at which Krylov becomes more efficient than R-MJP depends on the matrix's sparcity structure and the largest element of the matrix  $\max\limits_{n}{|a_{nn}|}$ (related to the effective time step in each of the methods). Moreover, in Sec. \ref{sec:benchmarking}, we focused on the time scaling of each method which, when implemented correctly, is independent of the hardware. However for the examples under investigation, using our code\cite{github} and hardware (described above) the critical size at which Krylov becomes more efficient is when the R-MJP and Krylov lines in Fig. \ref{fig:scaling} intercept.
}

\subsubsection{J-MJP is inefficient}\label{sec:J-MJPac}

As mentioned in Sec. \ref{sec:Bayes} the samplers generated by \#3, \#4, and \#5 obtain samples from the full posterior $p(\theta|\bar{s})$.  Consequently, the samplers from Methods \#3, \#4, and \#5 recover nearly identical probability distributions over model parameters $\theta$.

Meanwhile, Method \#2 uses the set of number of virtual jumps $\bar{k}$ as a latent variable. To effectively compare the performance of J-MJP (Method \#2) to other methods, it is essential to look not only at execution time but also to assess how well an equivalent number of samples describes the posterior.  
This becomes especially important since sampling $\bar{k}$ introduces extra variables equal to the number of observations. This inclusion dramatically increases the dimensionality of the sample space. With this increased dimensionality,  more samples may be required to properly characterize the joint posterior between $\theta$ and $\bar{k}$.

\begin{figure}%[H]
    \centering 
    \includegraphics[width=\textwidth]{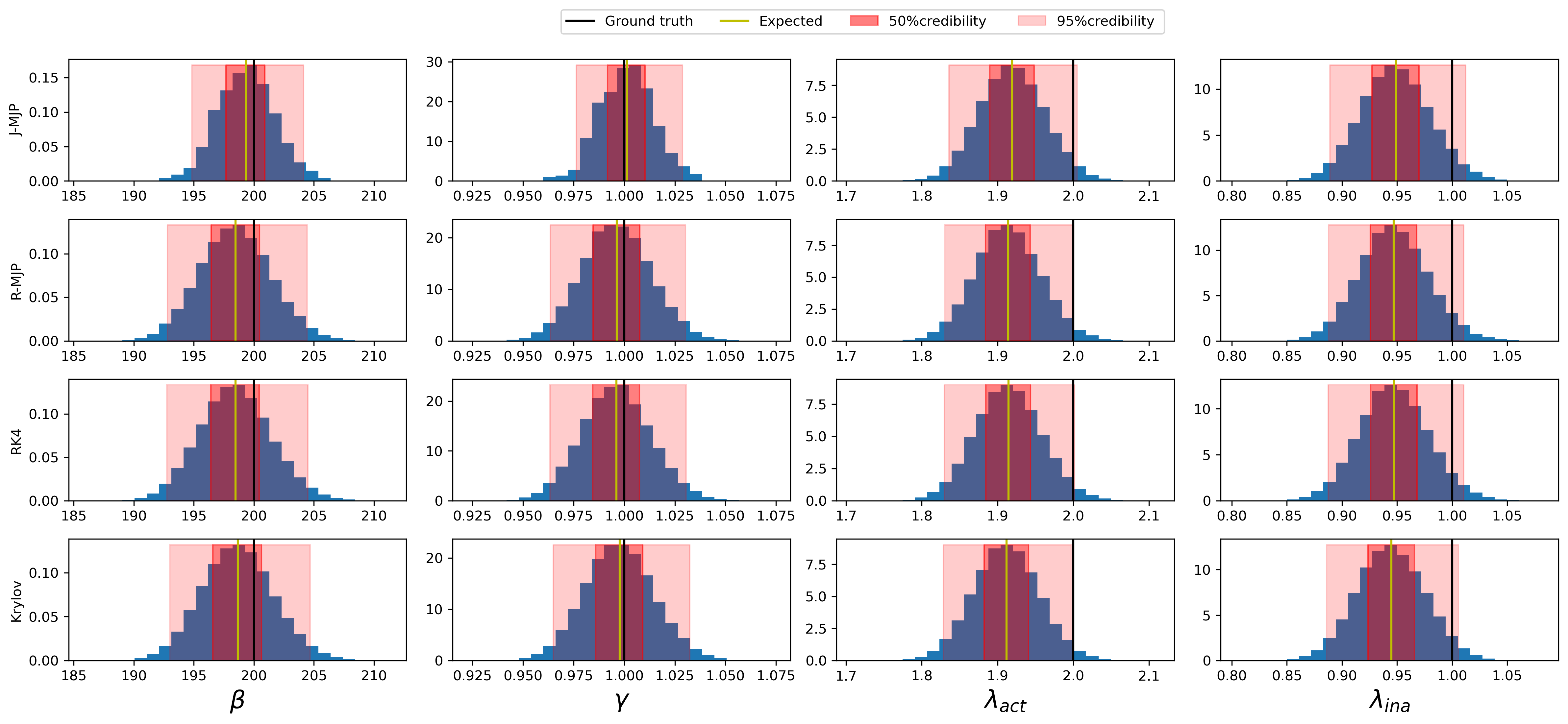}
    \vspace{-1.cm}
    \caption{ \change{Histogram of the posterior samples of each kinetic parameter for the two state birth-death. The histogram samples are compared with the simulation's ground truth (parameter values used for generating synthetic data). In each of these histograms we highlight the estimated expected value of the posterior (corresponding to the sample mean of each parameter) as well as the estimated $50\%$ and $95\%$ credible intervals (corresponding to the samples' 25th to 75th percentiles and 2.5th to 97.5th percentiles, respectively). We notice that} R-MJP, RK4, and Krylov yield coherent posteriors. \change{On the other hand} J-MJP, while placing the bulk of its probability density in a similar parameter region, recovers confidence intervals inconsistent with the other methods.}
    \label{fig:posterior}
\end{figure}

In Fig. \ref{fig:posterior}, the posteriors obtained by J-MJP and the other three benchmarked methods are shown to give inconsistent results in both learned parameters and their uncertainty (depicted as credible intervals in~\ref{fig:posterior}). 
This comparison was done for the two state birth-death example (Sec. \ref{sec:Gene_example}) with ground truth kinetic parameters $\theta = \{\beta, \gamma, \lambda_\text{act} ,\lambda_\text{ina} \} = \{200, 1, 2, 1\}$.  More examples in the Supplemental Information \ref{SIsec:cons}.

\begin{figure}
    \centering
    \includegraphics[width=.85\textwidth]{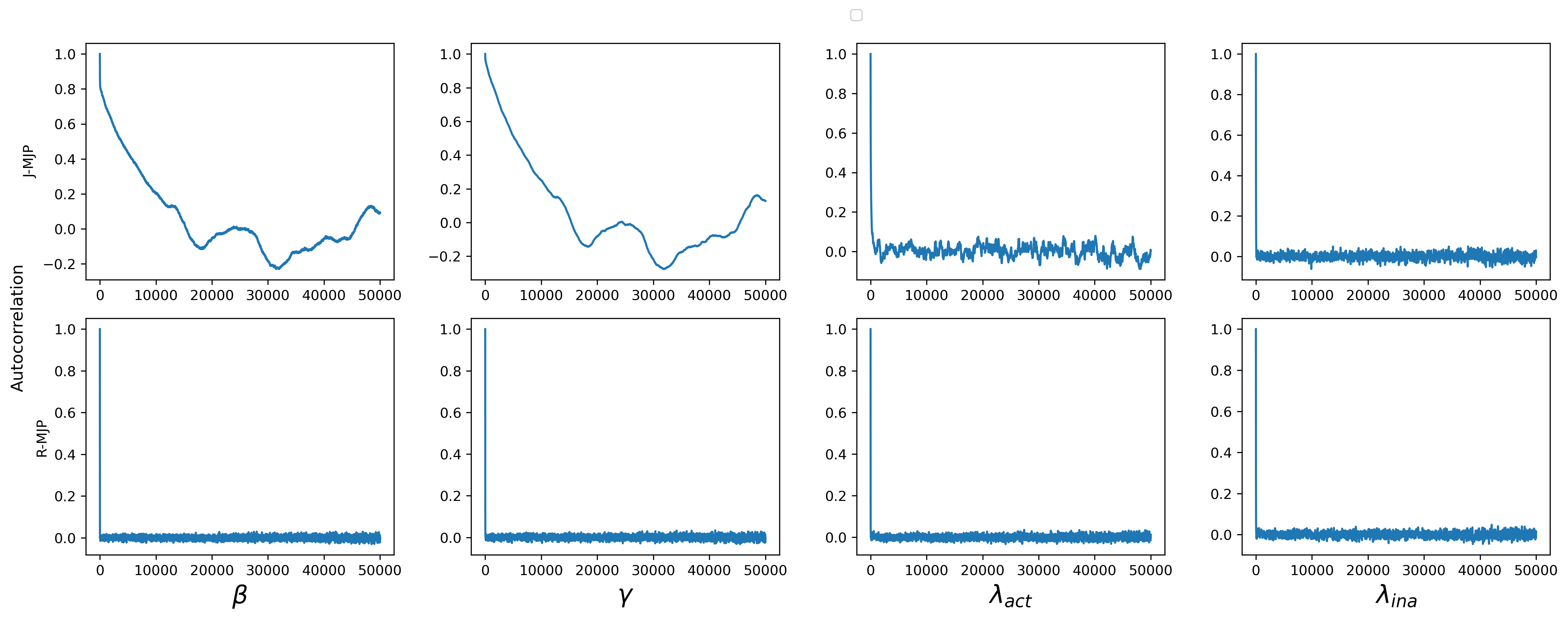}
    \vspace{-.5cm}
    \caption{Auto-correlation function \eqref{autocorr} for the kinetic parameters \change{sampled using J-MJP (on top) and R-MJP (in the bottom) within the two state birth-death example}. The results confirm that the sampler generated using J-MJP is inefficient, as the autocorrelation does not move to near 0 even after tens of thousands samples, while in R-MJP  the autocorrelation falls to zero after just a few hundred samples \change{(as occurs with Krylov and RK4)}.}
    \label{fig:autocorr}
\end{figure}

In order to demonstrate that the observed discrepancy is not due to a mathematical inconsistency but rather a result of insufficient sampling, Fig. \ref{fig:autocorr} presents the auto-correlation of the MCMC samples, defined by
\begin{equation}\label{autocorr}
\text{ACF}(\tau) \doteq \frac{\frac{1}{H-\tau}\sum_{h=\tau+1}^{H} (\theta^a_h - \bar{\theta}^a) (\theta^a_{h-\tau} - \bar{\theta}^a)}{\frac{1}{H} \sum_{h=1}^{H} (\theta^a_h - \bar{\theta}^a)^2} \ ,
\end{equation}
with $\theta^a_h$ being the $a$-th parameter of the $h$-th sample and $\bar{\theta}^a = \sum\limits_{h=1}^{H} \theta^a_h$. In simpler terms, $\text{ACF}(\tau) $ quantifies the average correlation between a sample and another with another sample $\tau$ steps away within the Markov chain.  \change{Indeed, the MCMC structure is intrinsic to J-MJP (and T-MJP) while the other three methods (R-MJP, RK4, and Krylov) calculate the vector times matrix exponential \eqref{ME_solution}, both within an MCMC framework and independently.}

An ideal scenario would have $\text{ACF}$ rapidly dropping from one to zero, signifying optimal mixing. Nonetheless, Fig. \ref{fig:autocorr} reveals that while this is achieved by R-MJP \change{(as well as RK4 and Krylov),}  for J-MJP sampled kinetic parameters remain correlated, even following thousands of iterations.
Consequently, at least an order of magnitude more samples are needed to characterize the posterior as proficiently with J-MJP as with RK4, Krylov and . This happens as a consequence of J-MJP having to sample a considerable number of latent variables.

\section{Conclusion}\label{sec:conclusion}

\begin{figure}
    %\centering
    \includegraphics[width=\textwidth]{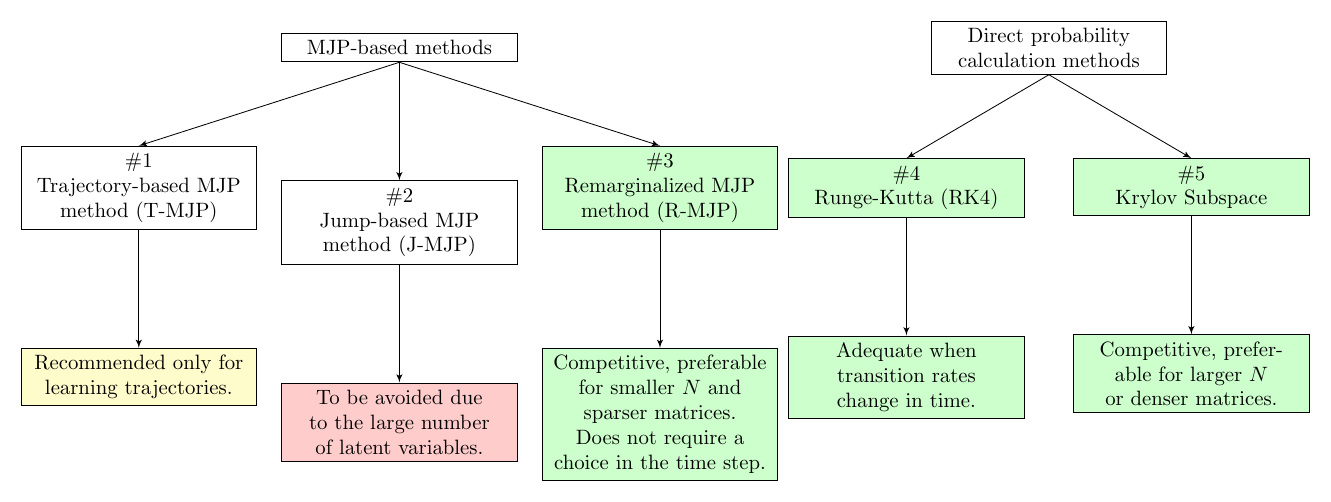}
    \caption{\change{Summary of recommendations for when to use each of method described in the present article based on the results reported in Sec. \ref{sec:benchmarking}.}}
    \label{fig:conclusion}
\end{figure}

Our comparison of different methods to avoid matrix exponentiation reveals that no one method is always preferable. Fig. \ref{fig:scaling} indicates that the most efficient method to generate sufficient samples of the posterior hinges on both the system and the state space: either the R-MJP (Method \#3) or the Krylov subspace (Method \#5), with the latter typically proving more efficient for significantly larger state spaces and the former seeming to have advantage for sparser matrices. However, the other methods should not be dismissed entirely for reasons \change{we discuss below.  Fig. \ref{fig:conclusion} summarizes the discussion by giving a recommendation for when to use each method. }

Although RK4 (Method \#4) performed significantly worse than all other methods, except the complete matrix exponential, it was the only method generalizable to cases where transition rates change in time -- such as induction~\cite{rahman2013single} -- since it doesn't invoke a time-constant generator matrix. 

Although we dismissed Method \#1 (T-MJP) for large state spaces due to its memory requirements, it is the only method allowing for trajectory inference and, as such, can still find important applications in some problems \cite{Kilic21generalizing}. 

Regarding Method \#2 (J-MJP), the result in Fig. \ref{fig:autocorr} indicates that it requires a much larger number of samples to characterize the posterior due to many latent variables introduced, thereby reducing the mixing for the kinetic parameters in $\theta$. The result is a sampler which mixed significantly worse than those using \#2, \#4, and \#5 which all marginalize over these latent variables.
 
Opting for synthetic data was crucial to demonstrate, in a controlled environment, the wall time scaling across state spaces size ranging in the hundreds to thousands of states. Our findings suggest that, for practical purposes, the choice between Krylov and R-MJP is highly contingent. \change{As such, we recommend} a benchmark \change{of these two methods} tailored to the specific dynamics under study \change{for anyone who needs their algorithm to run as efficiently as possible}.

Furthermore, although the time benchmark indicates that Krylov outperforms R-MJP in large state spaces, Krylov still demands more  memory allocation, associated with the choice of subspace size, $\kappa$. In contrast, \change{and this is absolutely critical to emphasize,} R-MJP does not require any such choices. The effective time step, $1/\Omega_\theta$, is determined by the system, and the selection of $k^\ast$ establishes a directly computed (and optimistically presumed) upper bound on the error. 

\section*{Acknowledgments}
We would like to thank Ioannis Sgouralis, Brian Munsky, Tushar Modi, Julio Candanedo, and Lance W.Q. Xu for interesting discussions in the development of the present article. SP acknowledges support from the NIH (Grant No. R01GM134426, R01GM130745, and R35GM148237).

\bibliographystyle{elsarticle-num} 
\bibliography{refs}

\newpage

\appendix
%\titleformat{\section}{\normalfont\large\bfseries}{SI \Alph{section}}{1em}{} % Section title format

\counterwithin{figure}{section}
\renewcommand{\thefigure}{A\arabic{figure}}

\section*{Supplemental Information to ``Avoiding matrix exponentials for large transition rate matrices''}

Pedro Pessoa$^{1,2}$, Max Schweiger$^{1,2}$, Steve Press\'e$^{1,2,3}$

$^1$ Center for Biological Physics, Arizona State University 

$^2$ Department of Physics, Arizona State University 

$^3$ School of Molecular Sciences, Arizona State University  

\section{Sampling scheme for the models in examples}\label{SIsec: sampling}
In this Supplemental Information section, we describe how to sample the posterior the Bayesian posterior, as mentioned in Sec. \ref{sec:Bayes}. 
First we will give an overview of the specific strategy for the MCMC based sampler known as  Metropolis-Hastings (MH) and how it was applied to the examples mentioned in the main text.  
In Sec \ref{SIsec:adaptative} we describe the form of adaptive MH used. In Sec \ref{SIsec:J-MJP} we expand upon the extra work necessary to sample $\bar{k}$ in method \#2. 
Finally, in Sec. \ref{SIsec:details} we give some more detail for the prior, Monte Carlo chain initialization scheme, and the parameters used in the adaptive scheme.

\subsection{Metropolis-Hastings sampling scheme}\label{SIsec:MH}

In order to create a MH based Markov chain Monte Carlo sampler drawing $\theta$ from the posterior $\theta \sim p(\theta|\bar{s})$ \change{as presented in Sec. \ref{sec:Bayes}, although the method described here can be used for sampling from any distribution. First, we ought} to initialize a sequence with a value $\theta_0$ and obtain a sequence (or chain) of samples $\{\theta_0,\theta_1, \ldots \theta_H \}$. With each iteration, from $\theta_h$ to $\theta_{h+1}$ following the steps:
\begin{itemize}
    \item[--] Sample a candidate $\theta_{\text{prop}} $ from a proposal distribution $q$, $\theta_{\text{prop}} \sim q(\theta_{\text{prop}}| \theta_h)$, our choice of proposal distribution will be discussed later.
    \item[--] Calculate the acceptance probability $\rho$ as 
    \begin{equation} p_\text{acc} = \min \left(  \frac{p(\theta_{\text{prop}}|\bar{s})}{p(\theta_h|\bar{s})} \frac{q(\theta_h|\theta_{\text{prop}})}{q(\theta_{\text{prop}}|\theta_h)} ,1 \right) \ . \end{equation}
    \item [--] Accept the candidate with probability $p_\text{acc}$. If accepted, we have $\theta_{h+1} = \theta_{\text{prop}}$, otherwise $\theta_{h+1} = \theta_h$.
\end{itemize}
This ensures that the sample chain converges to the desired target distribution $p(\theta|\bar{s})$ \cite{Presse23}. In principle it is independent of the proposal distribution $q$ and the initialization $\theta_0$, nevertheless these should be carefully selected in order to have an efficient sampler --- meaning a quickly converging sampler exploring a significant area of the target distribution with as few samples as possible. 

A usual form of the proposal distribution when dealing with dynamical processes is to treat the set of kinetic parameters as a vector $\theta = (\theta^1,\theta^2,\ldots,\theta^\eta)$ and proposing from a multivariate normal distribution in log space
\begin{equation} \label{prop}
    q(\Psi_{\text{prop}}|\Psi_h) = \text{Normal}(\Psi_h,\mat{Z}) \ ,
\end{equation}
where $\Psi_h = (\Psi^a_h) = (\log \theta_h^a)$. With $\Psi_\text{prop}$ sampled from \eqref{prop}, we obtain $\theta^a_\text{prop} = e^{\Psi^a} $.
 The covariance matrix $\mat{Z}$ is initially set as $\mat{Z} = z \mat{I}$, where $z$ is a small positive value and $\mat{I}$ is the $\eta \times \eta$ identity matrix. For an efficient sampler for the data at hand, it is useful to find adapt $\mat Z$, meaning changing $\mat Z$ to values generating an efficient distribution. We explain how this is done in the following subsection.

\subsection{Adaptive Metropolis-Hastings}\label{SIsec:adaptative}

 The adaptation process of $\mat{Z}$ follows the following steps inspired by \cite{Haario01},

\begin{itemize}
    \item[--]  A set of $H$ vectors, $\{ \theta_1,\theta_2, \ldots, \theta_H \}$, is sampled using the Metropolis-Hastings algorithm with a proposal of the form \eqref{prop}.
    \item[--] Update $\mat{Z}$ using the last $H'$ elements

\begin{equation}\label{update_S}
\begin{split}
\mat{Z} &= \left(\frac{2.38}{\eta}\right)^2 \left[ \text{cov}( { \theta_{H-H'},\theta_{H-H'+1}, \ldots, \theta_{H} } ) + \varepsilon \mat{I} \right] \ , \\
\text{where} & \ \text{cov}( { \theta_1,\theta_2, \ldots, \theta_{H'} } ) = \left( \frac{1}{H'} \sum_{h=1}^{H'} \theta^T_h \theta_h \right) - \bar{\theta}^T \bar{\theta} \ .
\end{split}
\end{equation}
Here,  $\bar{\theta}$ represents the average of the samples $\bar{\theta} = \frac{1}{H'} \sum_{h=1}^{H'} \theta_H$, and $\varepsilon$ is a small arbitrary value.
\end{itemize}

This procedure is repeated several times\change{, more details as well as the choice of parameters in each example are found in Sec. \ref{SIsec:parameters},} following which the MH sample chain restarts using the final $\mat{Z}$ obtained from the adaptation process. Only the samples obtained after stabilizing $\mat{Z}$ are saved as samples from the posterior.

\subsection{Sampling jumps in J-MJP}\label{SIsec:J-MJP}

When sampling from method \#4 in \eqref{sample_J-MJP} we also apply a MH strategy.  When sampling $\theta$ from $p(\theta|\bar{s},\bar{k})$, we use the sampling scheme outlined in Sec. \ref{SIsec:MH}. When sampling $\bar{k}$  from $p(\bar{k}|\bar{s},\theta)$ we use a proposal for each $k^i$ separately as
\begin{equation}
    q(k^i_{\text{prop}}|k^i_h) = P_\text{Poisson}(k^i_{\text{prop}}|\Omega_\theta T^m) \ 
\end{equation}
which is identical to $p(k^i|\theta)$, see \eqref{final_cond}. This simplifies the acceptance probability as
\begin{equation}
\begin{split}
    p_\text{acc}^m 
    &= \min \left(  \frac{p(k^i_{\text{prop}}|\bar{s},\theta)}{p(k^i_h|s^i,\theta))} \frac{q(k^i_h|k^i_{\text{prop}})}{q(k^i_{\text{prop}}|k^i_h)} ,1 \right) \\    
    & = \min \left( \frac{ p(s^i|k^i_{\text{prop}},\theta)}{p(s^i|k^i_h,\theta))}  ,1 \right) \\ 
\end{split}
\end{equation}
where in the second line above we substitute \eqref{sam_k} and $p(s^i|k^i_{\text{prop}})$ is calculated using Algorithm \ref{alg:J-MJP}.

\subsection{Bayesian sampler details}\label{SIsec:details}
To construct the posterior sampler, as outlined in the preceding subsections, certain hyperparameters remain unspecified. In this section, we will detail the hyperparameters employed both for the sampler discussed in the main text and for those covered in the following Supplemental Information section.

\subsubsection{Prior}
To compute the Bayesian posterior $p(\theta|\bar{s})$ as given by Equation \eqref{Bayes}, we require two components: the prior $p(\theta)$ and the likelihood $p(\bar{s}|\theta)$.  While the discussion in Sec. \ref{sec:inference} presents the likelihood, the choice of prior is tailored to the specific problem under consideration. Since our objective here was to characterize the different methods, we opt to use a prior as uniform as possible while keeping an analytic function for safe implementation within the multiple algorithms.  Thus for all inference processes studied here we choose {a normal prior for the logarithm of the kinetic parameters, $\Psi$ as defined in the paragraph containing equation \eqref{prop},}
\begin{equation}
    \label{prior}
    p(\Psi) = \prod_{a=1}^\eta \frac{1}{\sigma\sqrt{2\pi}} \exp\left( -\frac{(\psi^a - \mu)^2}{2\sigma^2} \right) \ .
\end{equation}
with $\mu = 1$ and $\sigma = 10^5$. 
Given that $ \sigma $ is significantly large, the prior exerts only a minimal influence on each individual kinetic parameter. Specifically, for $ \theta^a $ values ranging from $ 0.01 $, the smallest ground-truth kinetic parameter in our study, to $ 800 $, the largest ground-truth value, the prior remains roughly uniform up to four decimal places.

\subsubsection{Initialization}

Initializing the sampler, or selecting the initial value $\theta_0$, entails selecting a value of $\theta$ that can rapidly converge to regions of high posterior probability while avoiding numerical instability. Otherwise it is worth mentioning that the initialization is not critical. Given enough time, the sampler will explore the entirety of the parameter space, converging to the true posterior distribution.

For the birth-death process, Sec. \ref{sec:Birth-Death}, the initialization is done by choosing a naive estimator of the kinetic parameters. This means selecting $\theta_0 = \{\beta_0,\gamma_0\}$ as the least square errors from the mass action to the data, meaning
\begin{equation}
     \{\beta_0,\gamma_0\} = \arg \min_{\beta, \gamma} \left[\sum_i s^i - \frac{\beta}{\gamma} \left(1-e^{-\gamma T^i} \right) \right]^2 \ .
\end{equation}

While in the previous example mass-action is useful, for the two state birth-death system, Sec. \ref{sec:Gene_example}, the mass action is less informative as the observations exhibit larger variance. Hence, the initialization will use more arbitrary values, in this case by using the initial parameters $ \theta_0 = \{ \beta_0, \gamma_0, \lambda_{\text{act} 0}, \lambda_{\text{ina} 0} \} $ set as $ \beta_0 = \max_i s^i $ and $ \gamma_0 = \lambda_{\text{act} 0} = \lambda_{\text{ina} 0} = 1.5 $. While these starting values lie outside the regions of high posterior density, the sampler effectively guides them towards such regions in a reasonable timeframe. 

Finally, for the autoregulating gene, Sec. \ref{sec:autorel}, we initialize with random values within the same order of magnitude of the expected ground truth, meaning $
\theta_0 = \{ \beta_{R 0}, \beta_{P 0}, \gamma_{R 0}, \gamma_{P 0}, \lambda_{\text{act} 0}, \lambda_{\text{ina} 0} \} $ with 
$ \beta_{R 0} = 10 u^1$,
$ \beta_{P 0} = .5 u^2$,
$ \gamma_{R 0} = .1 u^3$,
$ \gamma_{P 0} = .05 u^4$,
$\lambda_{\text{act} 0} = 1 u^5$, and
$ \lambda_{\text{ina} 0} = .1 u^6 $ and $u^i \sim \text{Uniform}(.5,1.5) \  \forall i$.

\subsubsection{Parameters in the adaptive scheme}\label{SIsec:parameters}

The important parameters for the adaptive scheme are: the initial covariance matrix for the proposal distribution, for which we use $\mat{Z} = z I$ with $z = 10^{-8}$; the $\varepsilon$ used to update $\mat{Z}$ in \eqref{update_S}, for which we use $\varepsilon = 10^{-12}$; the number of steps between updating, for which we use $H=300$; and the number of samples used to update $\mat{Z}$ for which we use $H' =100$. Besides these parameters, it is also important to determine when to stop updating $\mat{Z}$.

In the case of the birth-death process, the initialization guarantees the sampler starts in a region of high posterior probability. Consequently, we terminate the updating procedure after completing 10 cycles, each consisting of 100 samples, where the acceptance rate \change{(}defined as the ratio of accepted proposals\change{)} falls within the range of $0.2$ to $0.5$. Note that the counting of cycles is independent of the updating of $\mat{Z}$ at each 300 samples, thus the sampler has to ``survive'' at least 3 changes in $\mat{Z}$ by keeping a reasonable acceptance rate before settling.

For the other two processes, the initialization starts in regions of low posterior probability. In these scenarios, the adaptive procedure first guides the sampler towards a region of higher posterior probability. This is done by performing the same adaptive process until completing 10 cycles where the acceptance is found between $0.2$ and $0.5$. Once this is achieved, we reset the adaptive process entirely, including the covariance matrix $\mathbf{Z} = zI$, but we initialize $\theta_0$ at the highest posterior value found by the samples up to that point. The process is again terminated after 10 cycles (20 cycles in the autoregulatory gene network example), each with 100 samples, when the acceptance rate settles between $0.2$ and $0.5$.

\clearpage
\newpage
\section{Consistency Results}\label{SIsec:cons}

\begin{figure}[p]
\begin{center}
   \includegraphics[width=.6\textwidth]{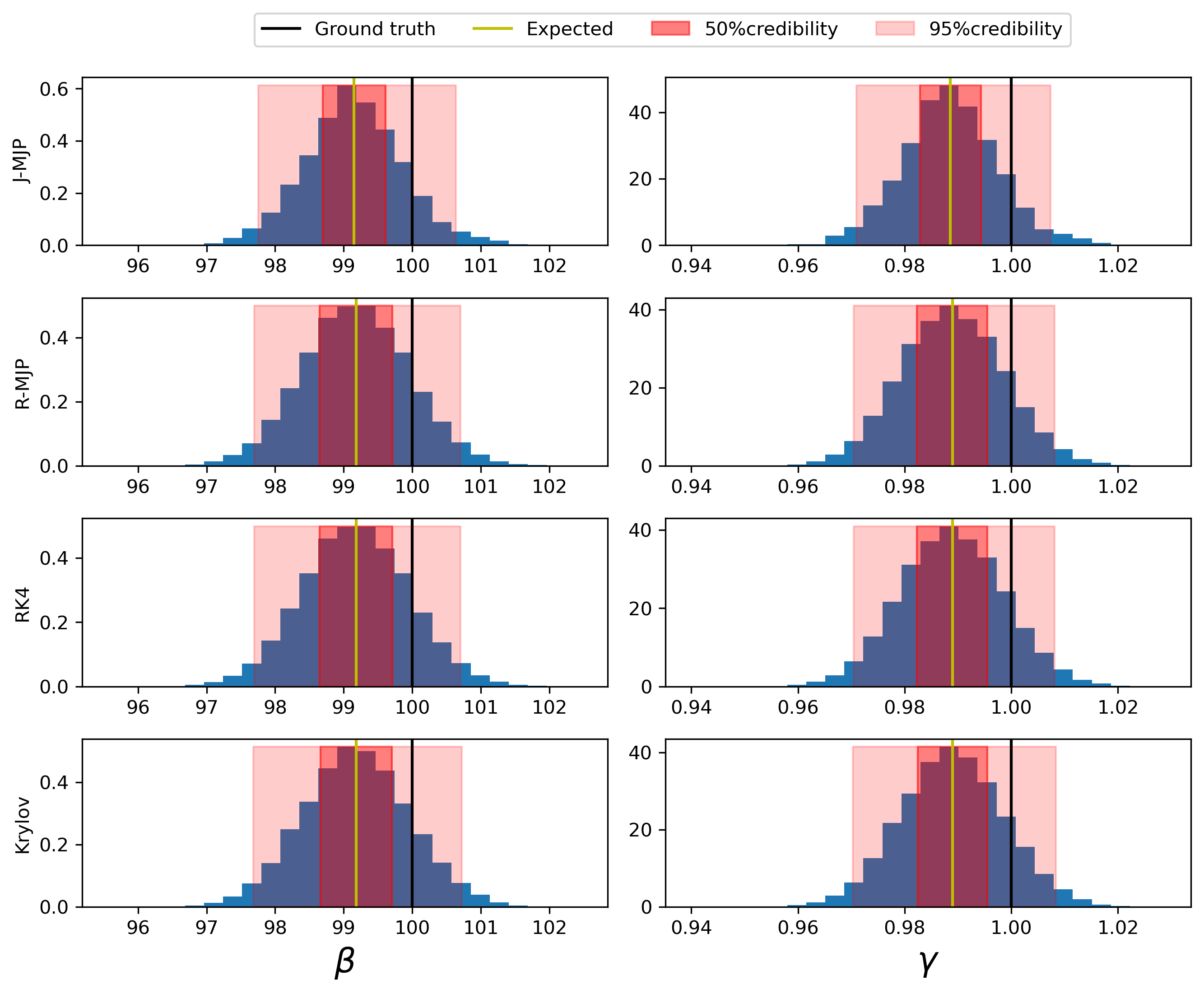}
   \includegraphics[width=.6\textwidth]{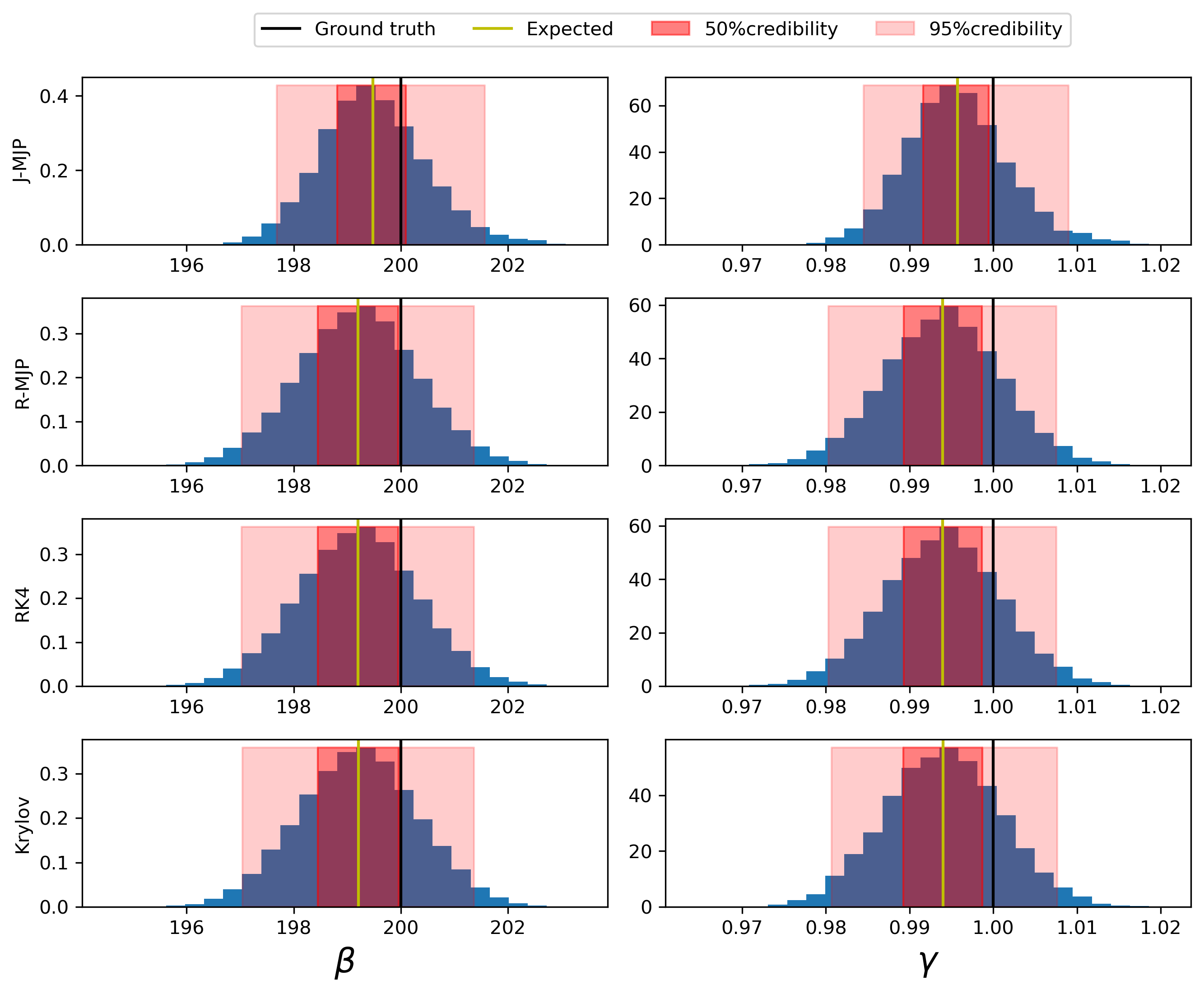}
\end{center}
\caption{ Density histogram of samples from the posterior distribution for the birth-death process (Sec. \ref{sec:Birth-Death}). Different panels represent  posterior obtained with datasets of varying kinetic parameter. \change{See Fig. \ref{fig:posterior} in the main text for conventions}.
}\label{fig:SIBD}
\end{figure}

\begin{figure}
\begin{center}
    \includegraphics[width=.9\textwidth]{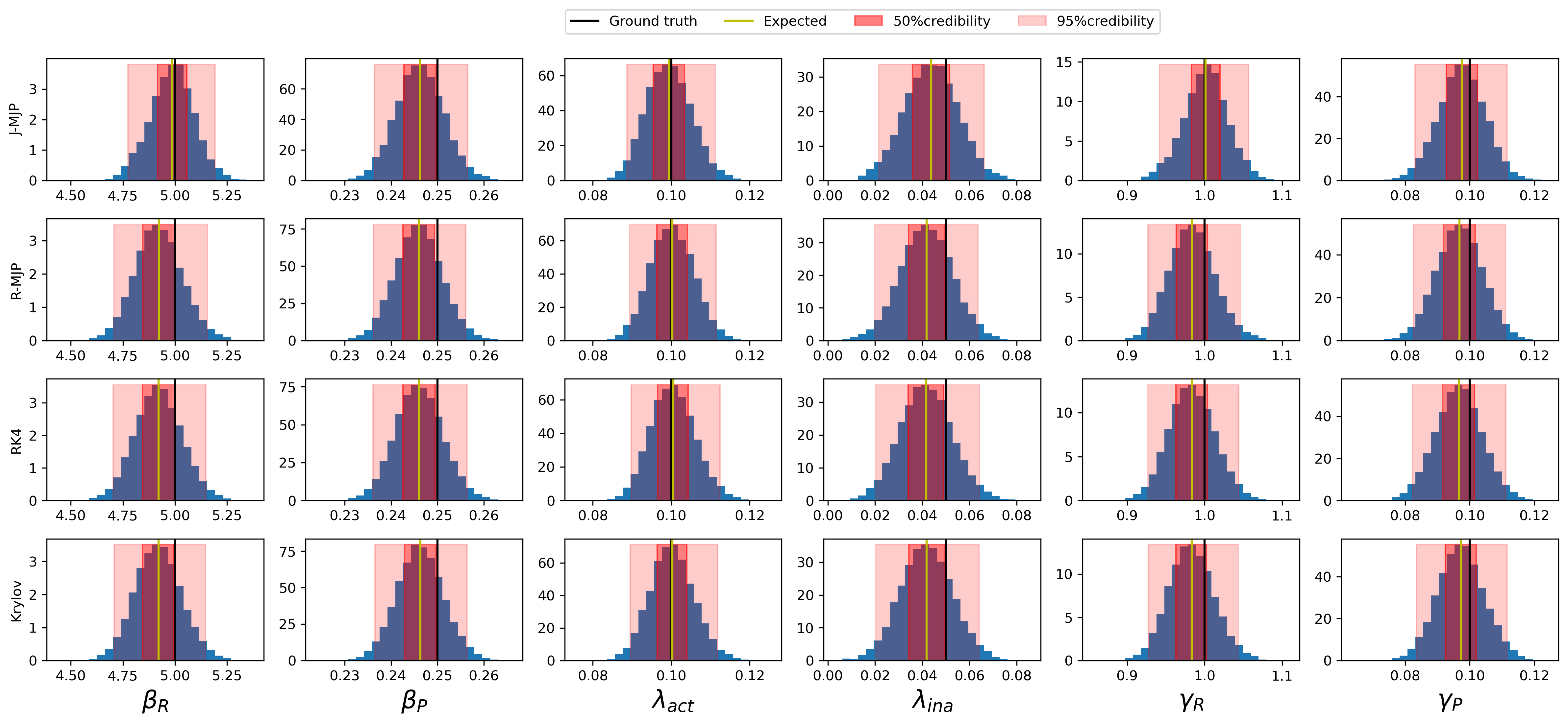}
    \includegraphics[width=.9\textwidth]{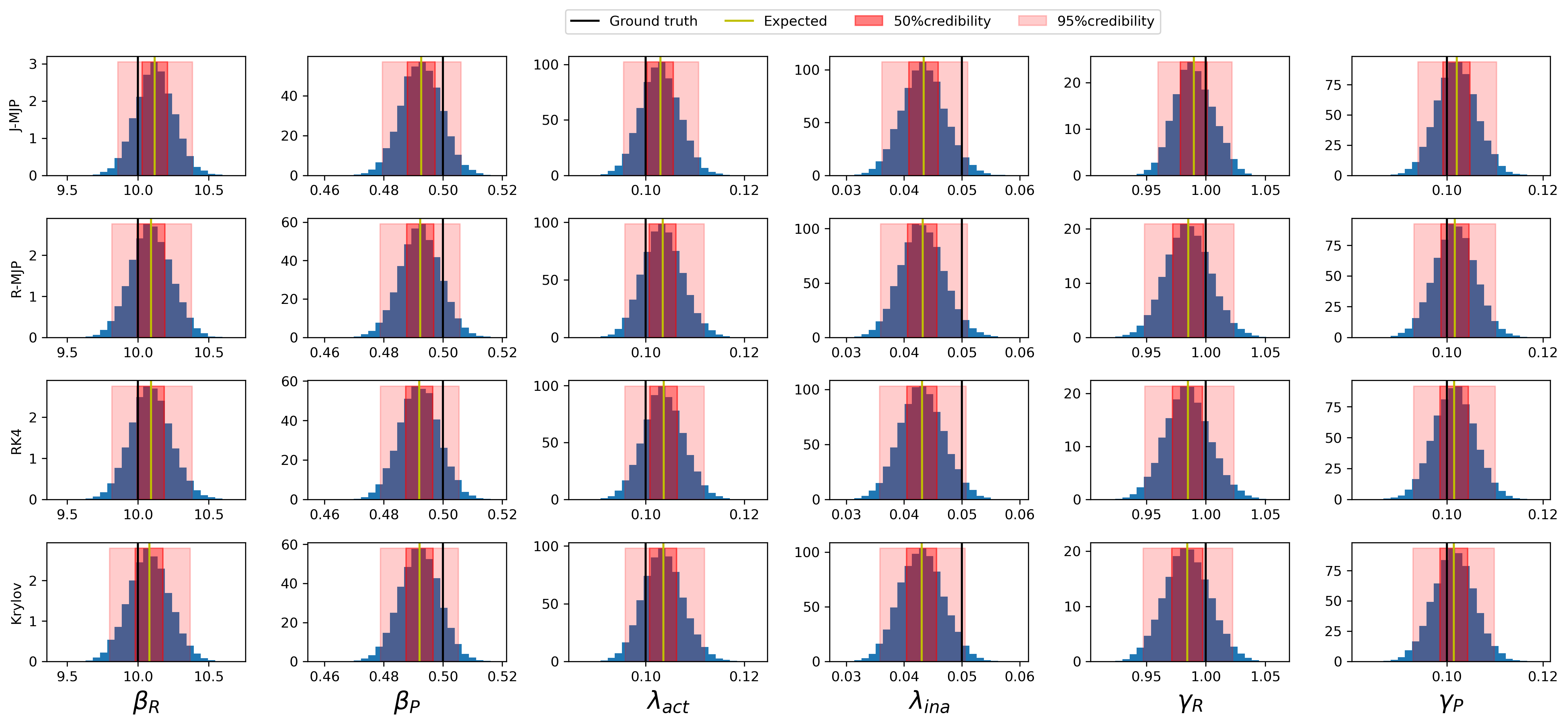}
\end{center}
\caption{ Density histogram of samples from the posterior distribution for  the autoregulating gene process (Sec. \ref{sec:autorel}). Different panels represent the posterior obtained with datasets of varying kinetic parameters.  \change{See Fig. \ref{fig:posterior} in the main text for conventions}.
}
\label{fig:SISTS}
\end{figure}

In this Supplemental Information section we present samples from the posterior for all kinetic parameter values treated in Sec. \ref{sec:inference} but not plotted in the main text. 
The posterior sample histograms for the birth-death process (Sec. \ref{sec:Birth-Death}) are found in Fig. \ref{fig:SIBD}, the ones for the two state birth-death (Sec. \ref{sec:Gene_example})
are found in Fig. \ref{fig:SI2S}, while the autoregulating gene example (Sec. \ref{sec:autorel}) is found in Fig. \ref{fig:SISTS}.

To ensure that one can obtain sufficient samples to fully characterize the posterior, $p(\theta|\bar{s})$ in feasible time we obtain $250000$ samples of the kinetic parameters, $\theta$, for each dataset analyzed. \change{In the J-MJP, we remove all other variables keeping only the samples of the kinetic parameters ( which is equivalent to marginalizing these latent variables). The density histograms are then constructed from these samples.} Each histogram focuses on individual elements of  $\theta$, \emph{e.g.}, for the birth-death process we have $\theta = \{\beta,\gamma\}$ and the density histogram for the sampled values of $\beta$ in an approximation of the marginalized posterior\change{,} $p(\beta|\bar{s}) = \int \dd \gamma \ p(\theta = \{\beta,\gamma\}|\bar{s}) $.  These histograms are juxtaposed with the sample average (approximating the marginal posterior expected value) and the 50 \% and 95\% credible intervals. See Fig. \ref{fig:posterior} in the main text for all conventions.

\begin{figure}[p]
\begin{center}
    \includegraphics[width=.8\textwidth]{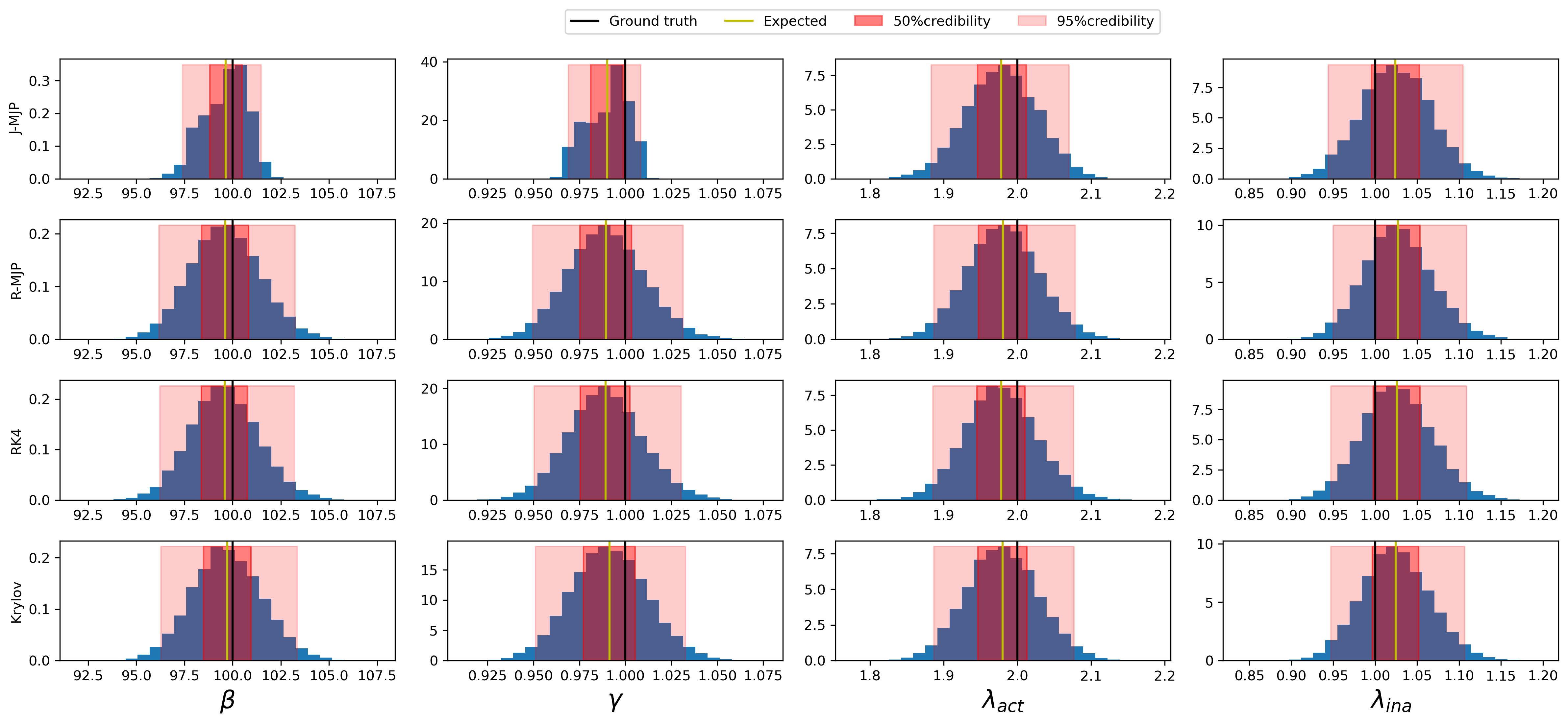}
    \includegraphics[width=.8\textwidth]{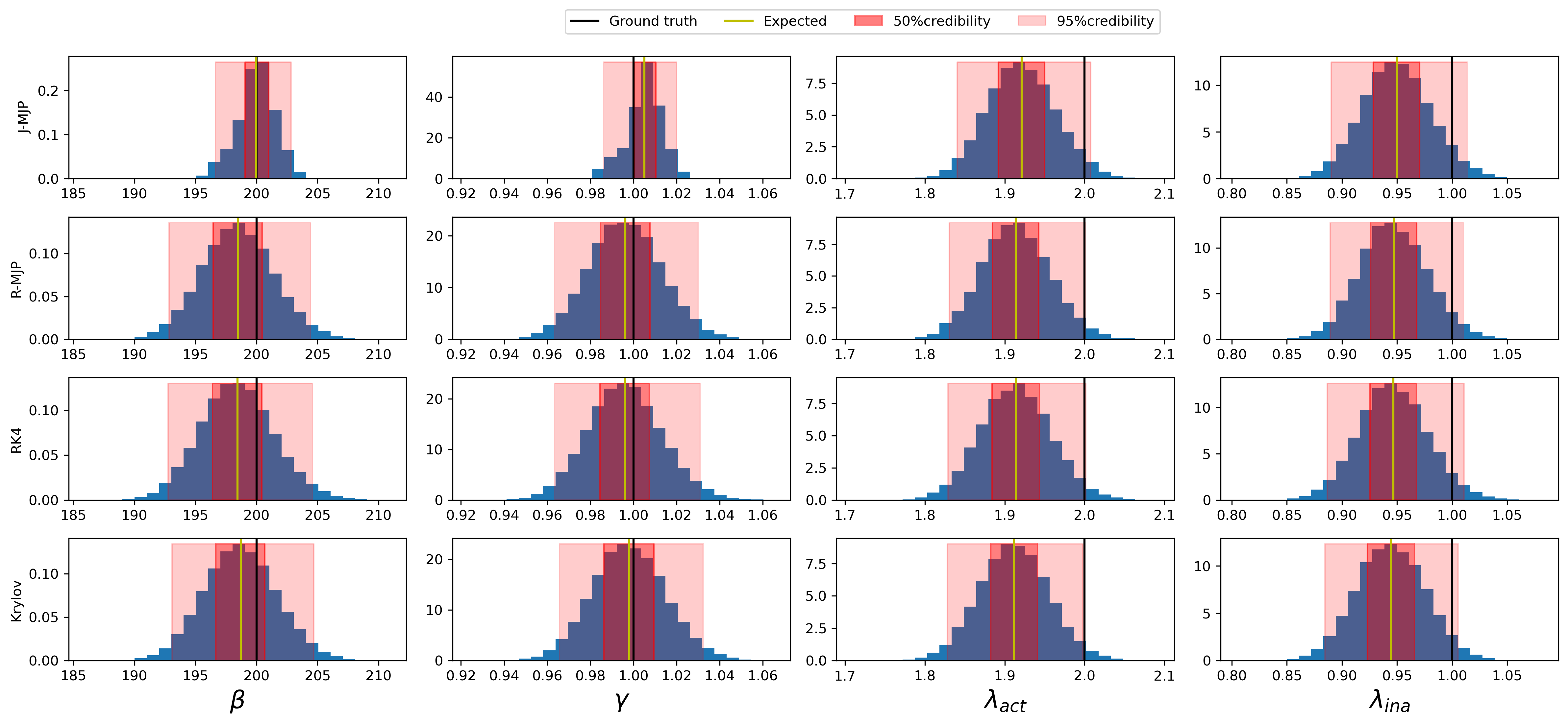}
\end{center}
\caption{ Density histogram of samples from the posterior distribution for the two state birth-death process (Sec. \ref{sec:Gene_example}). Different panels represent the posterior obtained with datasets of varying kinetic parameters. \change{See Fig. \ref{fig:posterior} in the main text for conventions}. }
\end{figure}

\end{document}